\documentclass[12pt,preprint]{aastex}
\newcommand{\lsp}{LS~I~+61$^{\circ}$303}
\newcommand{\grs}{GRS 1915+105~}
\newcommand{\lsi}{LS~I~+61$^{\circ}$303~}

\newcommand{\beq}{\begin{equation}}
\newcommand{\eneq}{\end{equation}}
\begin{document}

\title{Radio Spectral Index Analysis and Classes of Ejection
in  \object{\lsp}.}
   \author{M. Massi \and M. \,Kaufman Bernad\'o\altaffilmark{1}}
\affil{Max Planck Institut f\"ur Radioastronomie, Auf dem H\"ugel 69, 53121
Bonn, Germany}
\email{mmassi@mpifr-bonn.mpg.de}
\altaffiltext{1}{Humboldt Research Fellow}

\begin{abstract}
\lsi is a $\gamma$-ray  binary with periodic radio outbursts coincident with the orbital period of $P$=26.5~d. The origin of the radio emission is unclear,
 it could be  due either to a jet, as in microquasars, or  to the shock boundary between
the Be star and a possible pulsar wind.
We here analyze the radio spectral index 
over 6.7~yr from Green Bank Interferometer  data at 2.2~GHz and 8.3~GHz. 
We find two  new characteristics in the radio emission.
The first characteristic is that  the periodic outbursts indeed consist of two
consecutive outbursts;   the first
outburst is   optically thick, whereas the second outburst is
 optically thin.  
The spectrum of \lsi is  well reproduced by the shock-in-jet model
commonly used in the context of microquasars and AGNs: the optically thin spectrum
 is due to shocks caused by relativistic plasma (``transient jet")
traveling  through a pre-existing much slower steady flow  
(``steady jet"). This steady flow is 
responsible for the preceding  optically thick  spectrum.
The second characteristic we find  is  that the observed   
spectral evolution,  from  optically thick to optically thin emission,
occurs twice during the orbital period.
We observed this occurrence  
 at the orbital phase of the  main 26.5~d outburst 
and also at an earlier phase, shifted by  
$\Delta \Phi \sim$ 0.3 (i.e almost 8 days before).
We show that this  result qualitatively and quantitatively agrees
with  the   two-peak accretion/ejection model proposed in the past for \lsp.
We conclude that the radio emission in \lsi originates from a jet and  
suggest that the variable TeV emission   comes from 
the usual  Compton losses expected  as an important by-product 
in the shock-in-jet theory.
\end{abstract}



\section{Introduction}
The TeV-emitting source \lsi is a high-mass X-ray binary system,
where a compact object
travels through the dense equatorial wind of a Be star.
Due to the large uncertainty in the inclination of the system
($i=30^{\degr}\pm20^{\degr}$) the nature of the compact object has not yet been determined and either
a neutron star or a black hole are equally plausible candidates (Casares et al. 2005).
The most typical peculiarity of \lsi is a large periodic radio outburst \cite{taylor-gregory1982}
with a period of  $P_1$=26.496 d which is modulated by a period of  $P_2$=1667~d  (Gregory 2002).
The 26.5~d periodicity corresponds to the orbital period of the binary system \cite{hutchings-crampton1981}.
The origin of the second period is unclear. Observed also in the
H$\alpha$ emission line  \cite{zamanov-marti2000}, this long period
has been related to possible  variations in the equatorial wind
 of the Be star
(as a periodic shell ejection, \citealp{gregory-neish2002}).
However, the H$\alpha$ emission presents 
drastic variations in less than 24 h
more likely originating from a variable  source of ionization
(like a relativistic jet) than 
from improbably drastic changes in the density properties of the Be star disk
(Grundstrom et al. 2007).

\begin{figure*}[t!]
 \centering
\includegraphics[scale=0.6, angle=0.]{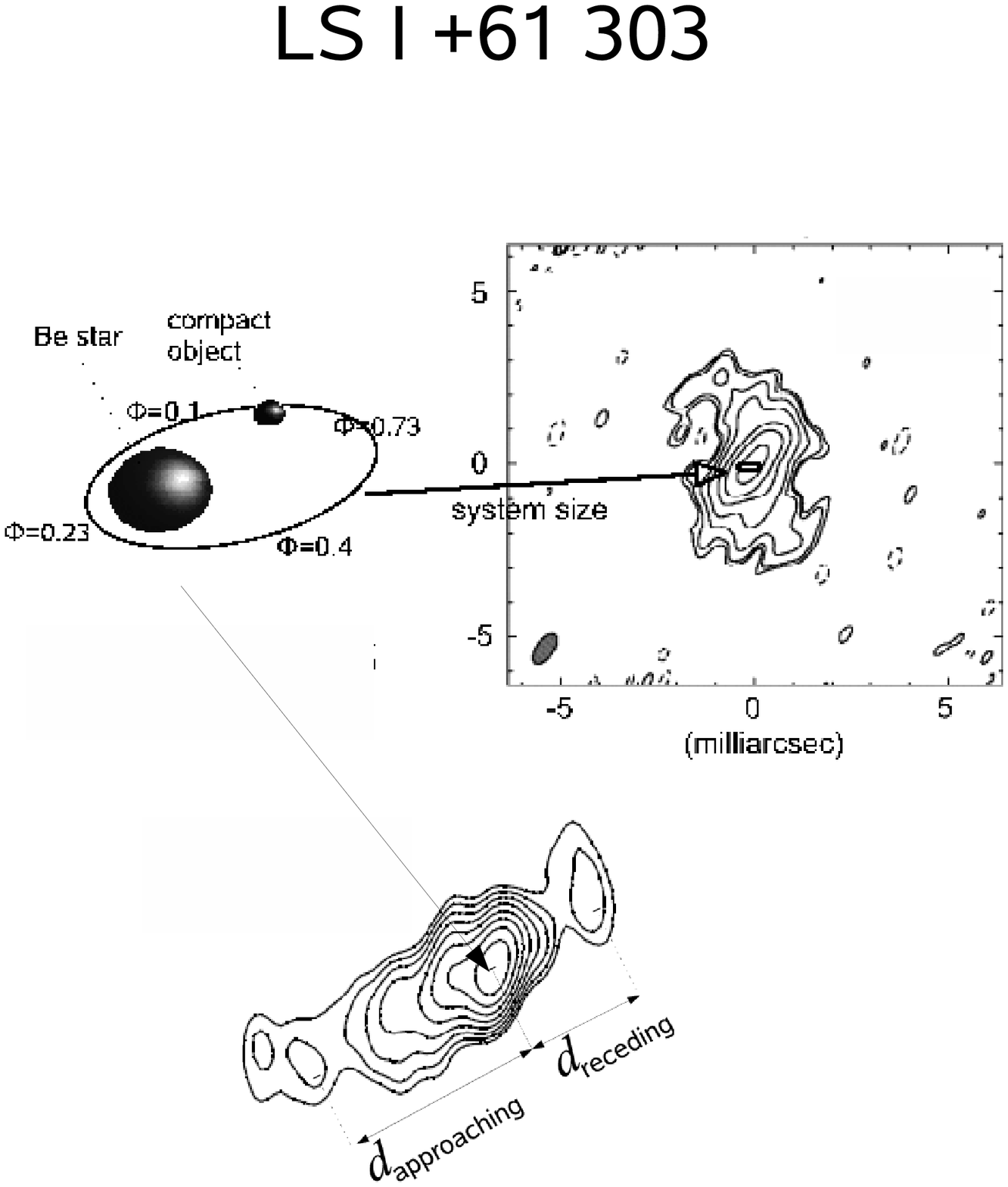}
   \caption{
Top-left: 
Sketch of the binary system.
Top-right: Radio observation by Taylor et al. (2000). 
The resolution of  the radio image is $\sim$ 0.95 mas $\times$ 0.45 mas and is
shown in the left bottom corner as a grey ellipse. 
The  dimension of the orbit is drawn in the center of the radio source.
Bottom: 
VLBA image by Dhawan et al. (2006).
The resolution of  the radio image is  $\sim$ 1.5 mas  $\times$ 1.1 mas, 
 therefore, as the previous image,  insufficient to distinguish any 
variations at the much smaller orbital scale. 
The image  can be interpreted as    
a steady jet (Sec. 4.2) with asymmetries due to  relativistic Doppler effects,
i.e.
with the receding jet  shorter than the approaching one. Consistently, also
the approaching and receding plasmoids at both sides 
are at different distances ($d_a \ne d_r$).
}
\label{radiomaps}
\end{figure*}

High resolution radio images of \lsi have generated controversial interpretations.
The bottom image in Fig.~\ref{radiomaps} shows an elongated structure, which is assumed to be  
 the cometary tail predicted at the shock boundary   between  the wind from the Be star
and the relativistic wind of a pulsar (Dhawan et al. 2006).
Alternatively, in the context of the microquasar model,
 there are two extreme substructures in the same image of Fig.~\ref{radiomaps} which can be interpreted
as two ejections around a re-formed steady jet. 
The different distances of these substructures from the
core as well as the assymmetry seen in the central elongated feature
 could be taken as the evidence
for Doppler boosting effects in the jet (Massi 2007 and Sec. 4.2).
Applying the two models, structures  in different images appearing at different position angles
are taken as a cometary tail always pointed away from the Be star (Dhawan et al. 2006) or
 as a precessing jet (Massi 2007 and references therein).
Taking into account the morphology of these high resolution VLBA images it is not possible to distinguish between
the two scenarios. Therefore a more decisive physical parameter is needed to decide between these two possible models.

The physical process producing the radio emission in both cases is the
synchrotron mechanism. However,
the evolution of the  spectrum corresponding to the  emission from
a  bubble of electrons in expansion 
is quite different from the evolution of a spectrum from
a relativistic jet with internal shocks 
as shown in the comparison of the bubble model  vs. the shock model 
by Hannikainen et al. (2006).
To distinguish between the microquasar and the pulsar model
 we therefore decided to study the spectral index accompanied by  
  an analysis of the flare amplitudes at different frequencies.
Sections~\ref{GBIdata} and \ref{results} we present 6.7 years of Green Bank Interferometer (GBI) data
and the results of our analysis.
In Sect.~\ref{classII} and Sect.~5 we show that the radio spectral index
is a very powerful tool that complements 
the  astrometry  and  X-ray observations.
Finally, in Sect.~\ref{conclusions}, we summarize our conclusions.

\section{Green Bank Interferometer Radio Data} \label{GBIdata}

\begin{figure*}[t!]
\begin{center}
\includegraphics[scale=0.3 , angle=-90.]{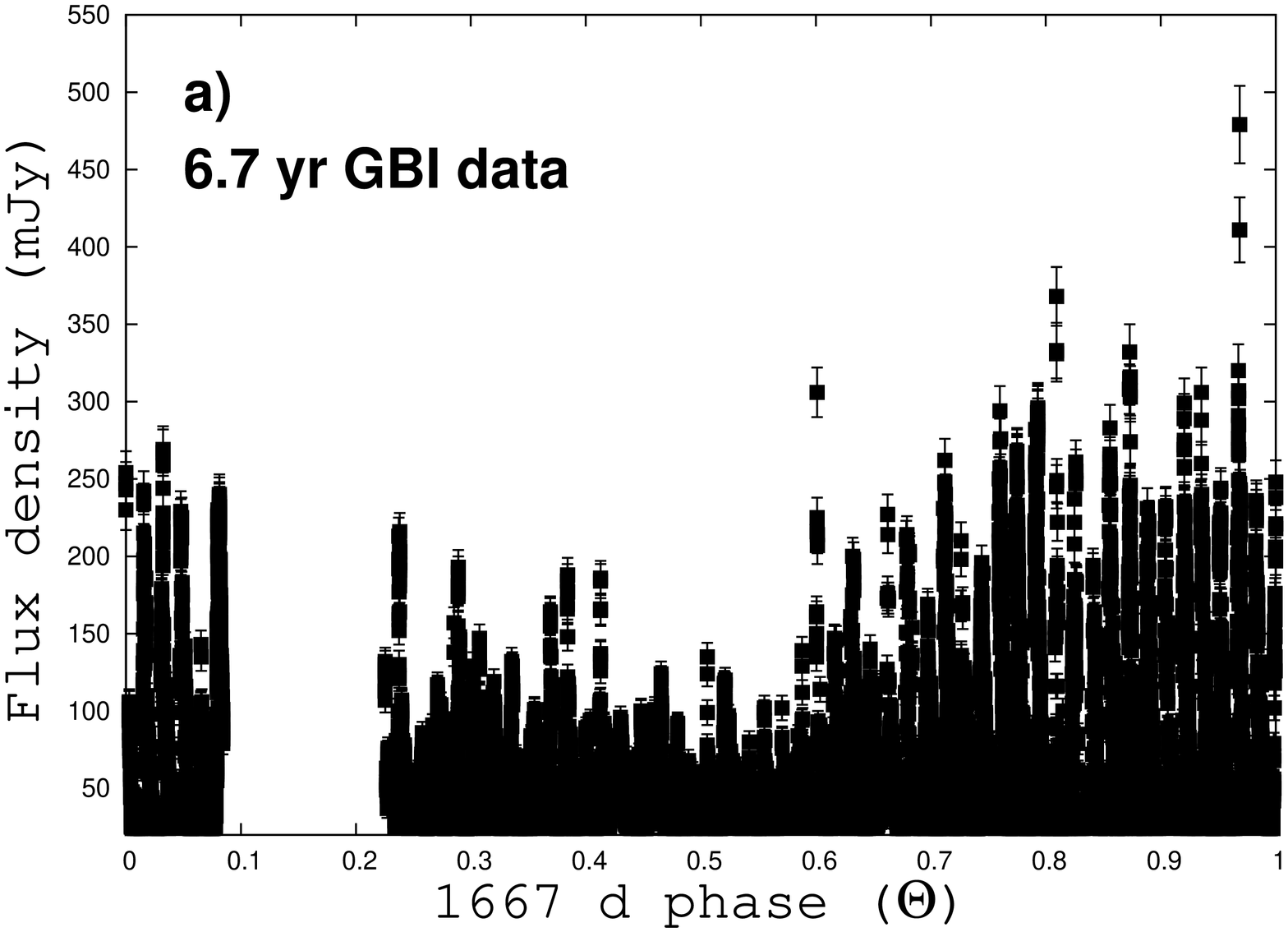}\hspace{0.5cm}
\includegraphics[scale=0.3, angle=-90.]{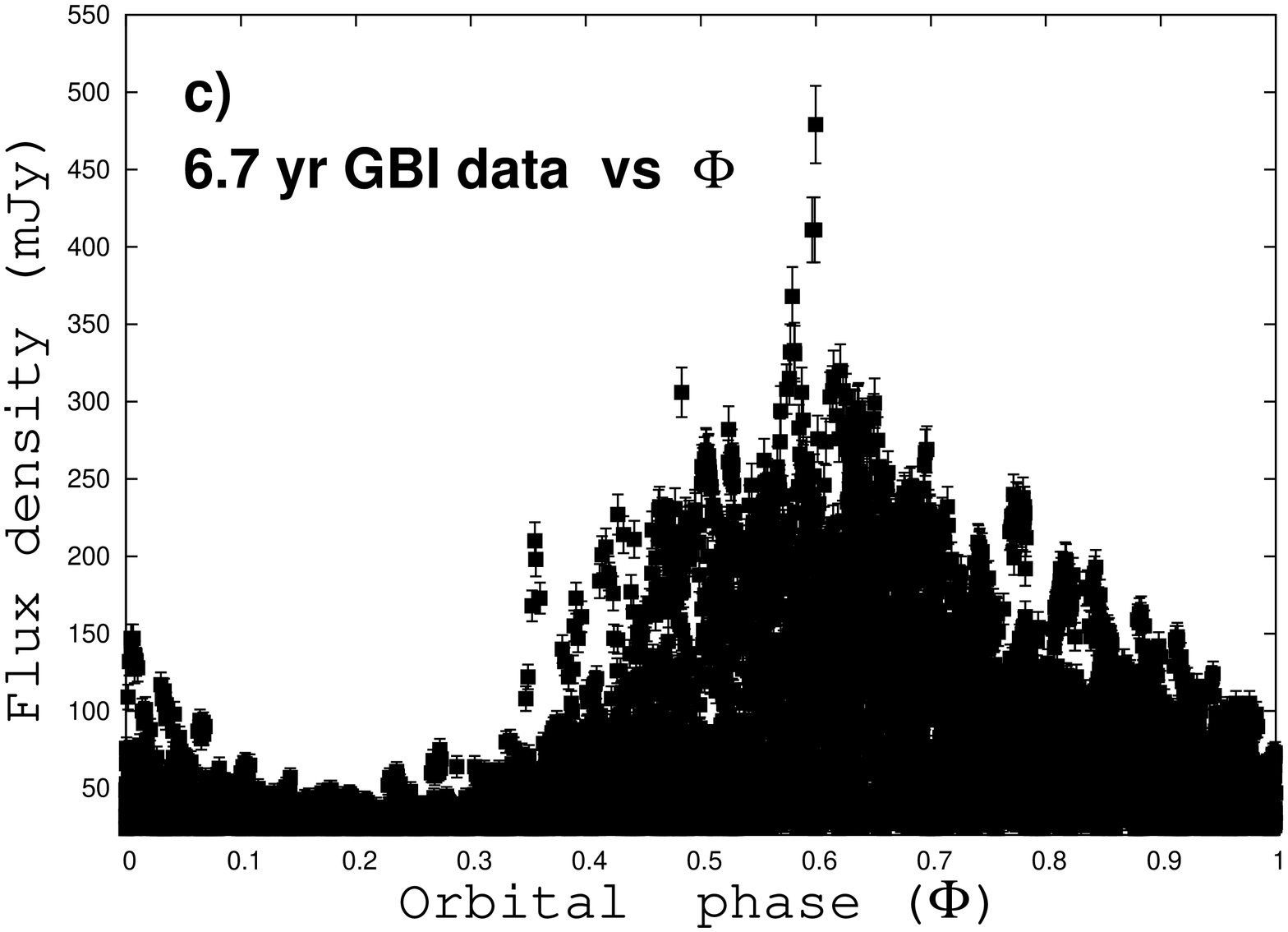}\vspace{0.5cm}
\includegraphics[scale=0.3, angle=-90.]{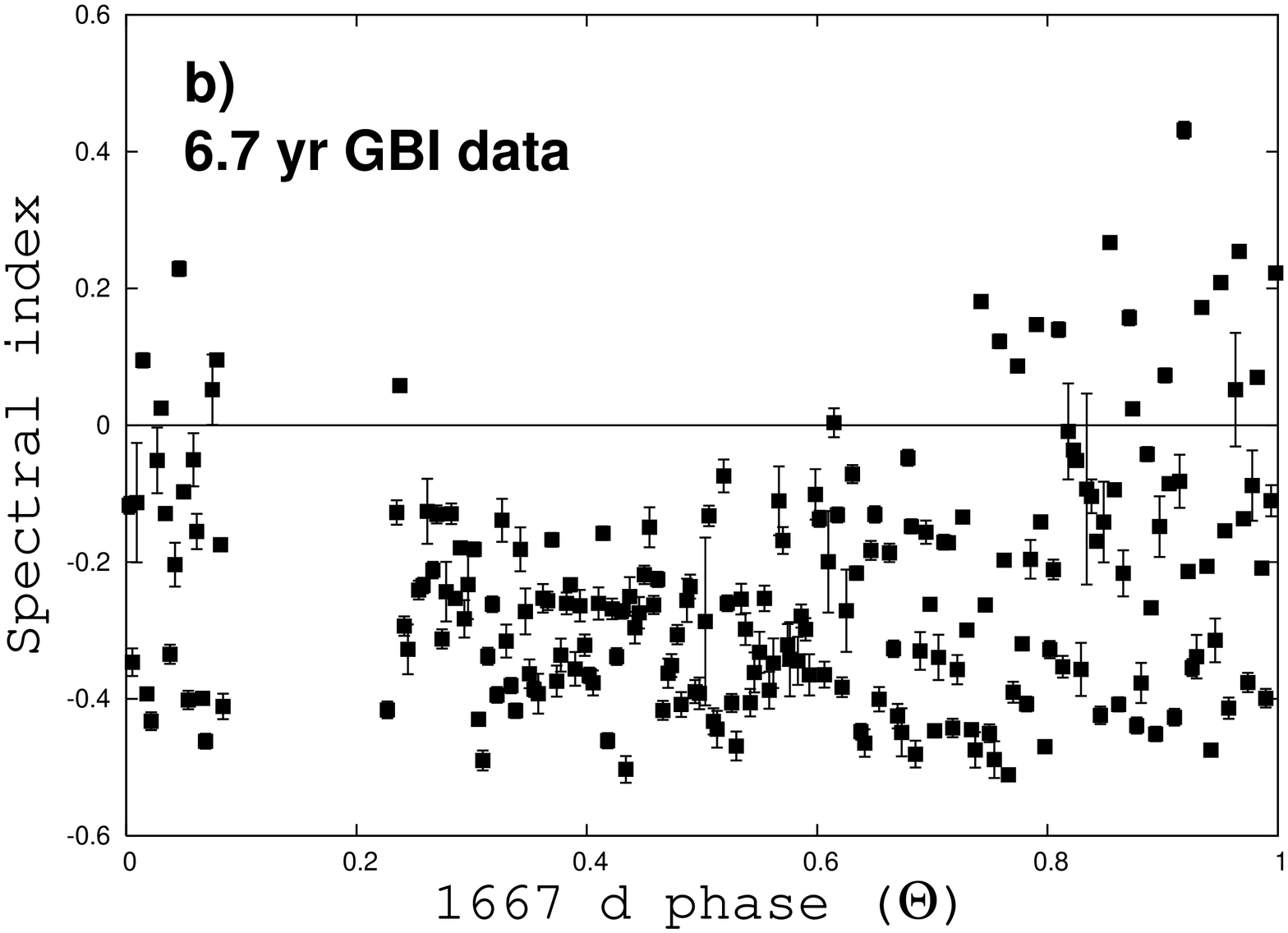}\hspace{0.5cm}
\includegraphics[scale=0.3, angle=-90.]{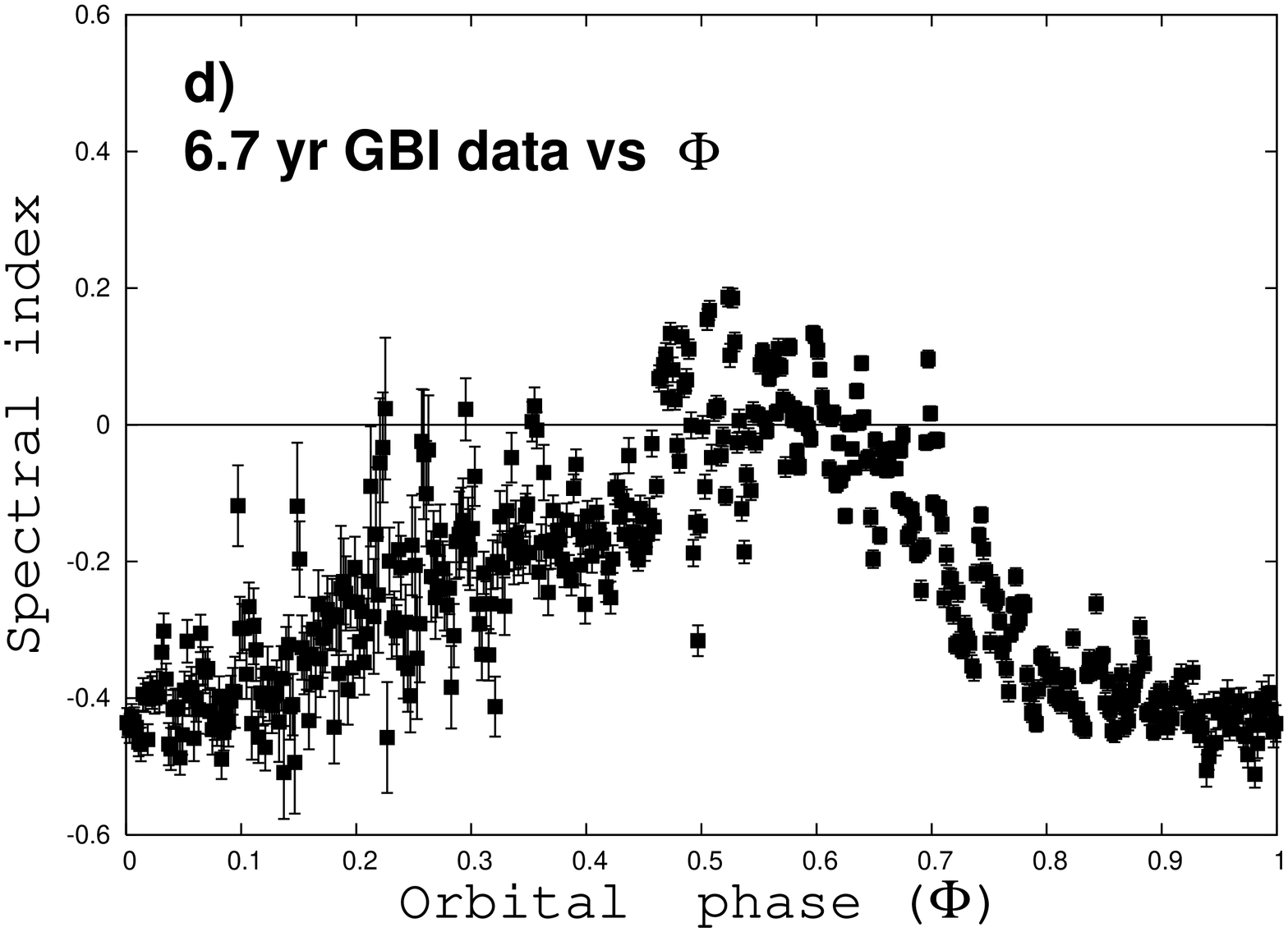}\vspace{0.5cm}
\end{center}
 \caption{
a: Radio light curve (8.3 GHz) vs $\Theta$, where 
$\Theta$ is related to ${(t-t_0)\over P_2}$ with  $t_0$=JD2443366.775 and 
period $P_2$=1667 d (Gregory 2002).
The maximum of the cycle is at $\Theta\sim 0.97$.
b: Spectral index vs $\Theta$. 
We averaged the folded spectral index curve binning the data into phase intervals
of $\Delta \Theta$=0.004 (equal to 7 d).
c: Radio light curve  vs $\Phi$,  the  orbital  phase, with 
$\Phi$  related to ${(t-t_0)\over P_1}$,
with $P_1$=26.496 d \cite{gregory2002}
d: Spectral index vs $\Phi$,
 averaged over $\Delta \Phi$=0.002,  ($\sim$ 1.3 h)}.
\label{GBI}
\end{figure*}

We here analyze 6.7 years of the  NASA/NRAO Green Bank Interferometer (GBI)
\lsi database at  $\nu_1=2.2$\,GHz and $\nu_2=8.3$\,GHz.
The database covers three periods: 49379.975$-$50174.710 MJD,  50410.044$-$51664.879 MJD and 51798.333$-$51823.441 MJD.
The samples of each period are continuous with at least two observations per day at each frequency.
Together with the flux densities at $\nu_1$ and $\nu_2$  and their corresponding errors, the GBI database also provides the spectral index,
 $\alpha={log(S_1/S_2)\over log(\nu_1/\nu_2)}$.
We calculated  the error $\Delta \alpha$, as 
$\Delta \alpha={0.434\over log (\nu_1/\nu_2)}\sqrt{( {\Delta S_1\over S_1})^2+
({\Delta S_2\over S_2})^2}$ and 
the weighted average $<\alpha>$ in each   phase bin  as
$\Sigma  {\alpha_i\over \Delta \alpha_i^2}\over \Sigma {1\over \Delta \alpha_i^2}$
with   
$\sigma=\sqrt{1\over \Sigma {1\over \Delta \alpha_i^2}}$.
In Figs.~\ref{GBI}-a and \ref{GBI}-c we show the whole
 dataset at 8.3 GHz, first folded with the long term period of $P_2$=1667~d and then folded with the orbital period of $P_1$=26.496 d.
The phase
 $\Theta$ refers to the fractional part of ${(t-t_0)}\over P_2$,
 whereas  the orbital phase, $\Phi$ refers to the fractional part of
${(t-t_0)}\over P_1$,
with t$_0$=JD\,2443366.775 (Gregory 2002).

\section{Results} \label{results}
Figure~\ref{GBI}-b shows the evolution of the spectral index, $\alpha$, during the 1667~d cycle.
A remarkable difference is obvious: the negative values of $\alpha$, i.e. optically thin emission,
correspond to a minimum in the flux density,
 whereas towards the maximum of the cycle
 $\alpha$ can also be zero or positive, i.e. the spectrum  can be 
flat or inverted.

In Fig.~\ref{GBI}-d,  the  6.7~yr dataset is folded with $P_1$.
This figure reveals that the spectrum is flat or inverted in the interval $\Phi  \sim 0.22 - 0.35$ and more clearly in the interval $\Phi  \sim 0.46 - 0.7$.
At $\Phi\sim$ 0.7,  $\alpha$ declines very rapidly  toward a value of $\alpha \simeq -0.4$.
In the following subsections, we will show that this evident transition
 of the spectral index is
an important tool for distinguishing  between the 
microquasar and the pulsar model.

\subsection{Pulsar vs. Shock-in-Jet Model} \label{results1}

Two types of radio emission are observed from X-ray binaries with jets.
They have completely different characteristics and are associated with
different kinds of ejections.
One corresponds to a flat or inverted spectrum and the other to an optically thin spectrum (Fender et al. 2004).
The  flat or inverted spectrum   covers the whole radio band 
      and  has been established also at  millimeter and infrared wavelengths 
(Fender et al. 2000; Fender 2001; Russel et al. 2006).
When  this kind of radio emission is 
spatially resolved it 
appears as a continuous  jet,
the so-called ``steady jet". In contrast, the optically thin
spectra are associated with isolated components, so-called  ``knots" 
or ``plasmoids", which are moving at relativistic speed away from the binary.
 This kind of ejecta are also known as a ``transient jet" (Fender et al. 2004, 2006).

The most important point is that the two 
kinds of radio emission and their corresponding types of ejection are
 related to each other;
if the optically thin spectrum (the transient jet) appears,
then it always happens after  the flat/inverted one (the steady jet).
Fender et al. (2004) associate this change in the radio spectrum to the parallel change that is observed in the X-ray states of these sources if passing from
 the low/hard X-ray state to the steep power-law state. In such a passage 
  there is an  increase in the bulk Lorentz factor of the jet.
This increase gives rise to 
shocks where the new  highly-relativistic plasma  catches
up with the pre-existing slower-moving material of the steady jet.
Fender et al. (2004) extended the AGN-shock model
to all the X-ray binary-jet-systems 
(i.e. microquasars). That model was originally  derived
by Marscher $\&$ Gear (1985), then generalized by Valtaoja et al. (1992), and
introduced in the context of X-ray binaries for \grs by Kaiser et al. (2000).
This internal shock model predicts  a clear trend in flux variations depending on frequency, which will be described  below.

The transition between the two kinds of radio emission
(optically thick to  optically thin)
 therefore is an important characteristic of microquasar systems.
This feature is  clearly shown in our data presentation.
 Figure~\ref{oo4}
 shows the two light curves and spectral index of  data at  $\Theta= 0.0-0.1$ 
(1455 measurements at each frequency) 
 folded with the orbital period $P_1$.
That figure shows a main peak of $S_{8.3{\rm GHz}}=270 \pm 15$  mJy at $\Phi=0.69$.
We call this outburst  {\it Peak$_1$}, which is an optically thick 
outburst, as one can determine
from the clear positive spectral index  shown in the figure (Top).
At 2.2 GHz the figure shows that {\it Peak$_1$}  
is followed by a large  outburst at $\Phi=0.82$; we call this {\it Peak$_2$}, 
($S_{2.2{\rm GHz}}=299 \pm 6$ mJy).
 At 8.3 GHz {\it Peak$_2$} corresponds to a minor outburst resulting in a clear
  optically thin spectrum.
In Fig.~~\ref{oo4}-Top the spectral index declines 
 between the two peaks with an inversion 
of the spectrum from optically thick to optically thin at $\Phi\simeq$0.75.  
After {\it Peak$_2$} the spectral index remains rather constant at a value of
$\alpha\simeq $-0.4.
In Fig.~\ref{oo5}-Top we analyze individual
 light curves (a 26~d data set, 
 126 measurements at each frequency) selected at a different
phase of $P_2$ $\Theta$=0.76.
The curve at 2.2 GHz clearly shows the two consecutive peaks.
At $\Phi \simeq$ 0.50 we have {\it Peak$_1$} with $S_{8.3{\rm GHz}}=230\pm 13$ mJy
 and $S_{2.2{\rm GHz}}=214 \pm 5$ mJy.
At  $\Phi$=0.61 both frequencies have a minimum, and
at $\Phi \simeq $0.68 {\it Peak$_2$} follows with 
$S_{8.3{\rm GHz}}$=193 mJy and  
$S_{2.2{\rm GHz}}$=269 mJy.
The averaged  spectral index for  the interval $\Theta$=0.7-0.8, 
shown in  Fig.~\ref{oo} panel 7, again  clearly shows
the decline of $\alpha$ between the two outbursts ($\Phi$=0.5-0.7).
In Fig.~\ref{oo5}-Bottom  we analyze other individual
 light curves  at $\Theta$=0.85 showing a yet better case, with more displaced
{\it Peak$_1$} and {\it Peak$_2$} with    
an  intermediate, additional  optically thick peak.
In the same plot (Fig.~\ref{oo5}-Bottom) the  H$\alpha$ emission-line 
measurements by Grundstrom et al. (2007) are given.
The H$\alpha$ value, still high at $\Phi=0.71$, 
shows a dramatic decline at $\Phi=0.749$ (1 day later)
corresponding with the onset of the optically thin outburst 
(see discussion in Grundstrom et al. 2007).

In other words, the  light curves indicate that there is not only  
a single outburst appearing after some delay at different
frequencies, but that there are two distinct outbursts at both frequencies. 
The physics behind the two outbursts seems to be quite different. 
In the first outburst the spectrum is flat or inverted,
 whereas the second outburst is clearly dominated by the peak at 2.2 GHz (optically thin outburst).
The optically thin outburst corresponds to different
conditions, as its much higher amplitude, the spectral index, and the
 H$\alpha$ emission-line measurements indicate. 

Let us first examine the pulsar model.
The variation of  the spectrum from  inverted to optically thin
would correspond to an expanding synchrotron-emitting region in any model.
 In the pulsar scenario it would be the two colliding  winds of a pulsar and
 Be star.
We can assume that the  initial region,
sufficiently compact to be optically thick,  
afterwards expands and becomes optically thin.  However,
following  van der Laan (1966)  this model predicts 
that the self-absorption turnover peak frequency will move to lower frequencies,
and the peak flux density will diminish 
(Valtaoja et al. 1992; Hannikainen 2006).
In other words, this model predicts only one outburst at each frequency, 
moving  to lower frequencies with lower flux densities.
Considering that in fact there are two outbursts, 
the adiabatic expansion could explain the small delay observable 
between {\it Peak $_1$} at 8.3 GHz and 2.2 GHz, i.e. 
the optically thick outburst,
 but cannot  explain why the flux would rise again producing {\it Peak $_2$}.

On the other hand,  the two interacting winds of the pulsar model suggest 
a prolonged   injection of energetic particles, rather than a single event.
Connors et al. (2002) have
shown  that  a prolonged injection of energetic particles can 
maintain an outburst
optically thin, as  is observed  for the pulsar PSR B1259-63, 
which is in an orbit around a Be star.
In PSR B1259-63  the spectrum  is optically thin 
throughout the outburst as is demonstrated in Fig.~3 of   Connors et al. (2002). 
In  Fig. 2  of Paredes et al. (1991) is shown
that a prolonged injection of energetic particles,
can reproduce only the optically thin
outburst ({\it Peak $_2$}) of \lsp.
Therefore, the model with
prolonged injection of energetic particles cannot explain
why an  optically thick  
{\it Peak $_1$}  should be  present in \lsi before {\it Peak $_2$}. 

The complex peak sequence that  we found in \lsi finds a natural 
explanation in the disc-jet coupling model described by Fender et al. (2004):
first there is a continuous outflow with flat or inverted spectrum, 
then an event (probably {\it Peak $_1$}) 
triggers a shock in this  slow optically thick outflow 
(Fender et al. 2004). At this point  the young growing shock creates 
the  optically 
thin outburst ({\it Peak $_2$}) (Valtaoja et al. 1992; Hannikainen et al. 2006).
Like for the observations of the microquasar  GRO J1655-40  explained
with the shock-in-jet model (Hannikainen et al. 2006),
 in \lsi one observes 
the increase of the flare amplitude with decreasing frequency, 
resulting in an  optically thin flare.
The agreement between  \lsi, with spectral index $\alpha\simeq-0.4$,
 and GRO J1655-40 
 is impressive. 
In Hannikainen et al. (2006, Fig. 4),  the spectral
index, $\alpha$, during  the outburst and during the decay remains in the 
range $\alpha=$-0.6 to -0.4.

During the stage of  ``growth" of the shock, 
the dominant cooling mechanism is inverse Compton losses producing
gamma-rays (Marscher \& Gear 1985).
Inverse-Compton losses 
fall off  with the radius as the shock
expands and are superseded by synchrotron losses.
It is interesting to see that
this  second stage of the shock-in-jet model, where synchrotron losses dominate, is different in different sources.
In this phase, after the peak, either the flux density decay  begins like in the blazar 3C 279 (Lindford et al.  2006) and   \lsp,
or the flux density remains roughly constant in a ``flat" stage,
like in 3C 273 and the microquasar GRS 1915+105 (T\"urler
et al. 2000; 2004). 
The analogous behaviour of \lsi and  the blazar 3C 279 during the second stage
is worth of note considering 
the low inclination angle for the orbit of \lsp.  
Moreover,  in another blazar, BL Lac, Marscher et al. (2008) recently found 
 evidence  for a double consecutive optical flare.
The authors discuss that  the first flare  corresponds
to a disturbance due to  explosive activity  near the black hole,
then the disturbance forms a shock wave responsible for the second flare.
A similar  interpretation can be applied to  the  double radio flare
of \lsp,
suggesting that \lsi is a microquasar with a jet close to the line of sight,
 i.e.  a microblazar,  where explosive activity and the shock wave
 are seen almost face on.

\subsection{The Two-Peak Accretion/Ejection Model} \label{results2}

Figure~\ref{oo4}-Top shows that  
the evolution  from an optically thick to an optically thin spectrum 
occurs twice, giving  
the curve  $\alpha$ vs $\Phi$  a double-peaked  shape.
In Fig.~\ref{oo} we show the whole 6.7~yr data set, divided in $\Theta$
intervals.
The  double-peaked curve   is  confirmed 
in the intervals $\Theta=0.7-0.3$, which corresponds to 
the maximum of the 1667~d cycle. 
In this $\Theta$ interval,  the spectrum is flat or inverted ($\alpha\geq$0)
in the two orbital phase ranges 
0.22$\pm$0.08 to 0.33$\pm$0.12  and  0.50$\pm$0.09 to 0.70$\pm$0.13. 
Around  the minimum of the 1667~d cycle, at $\Theta=0.4-0.5$, 
these two intervals,  in which  the spectrum is  inverted, seem to be  reduced to only  a few points with $\alpha\ge$0.

Figure ~\ref{flal} shows the data of  Fig.~~\ref{oo4} (i.e. $\Theta=0.0-0.1$)
plotted as spectral index vs. flux density.
The change from  an inverted spectrum  to optically thin emission
around orbital phase $\Phi\sim 0.75$ occurs at high flux densities.
This change has been explained  in Sect. 3.1 due to  shock propagation  
throughout the pre-existing slow jet formed during the phase with an inverted 
spectrum.
In contrast, the previous  change  around orbital phase 
 $\Phi\sim 0.48$ occurs at low flux densities.
The similar evolution of the spectrum,  
flat or inverted spectrum (steady jet) 
to optically thin jet (transient  jet) indicates that
the same physical process does indeed occur twice in \lsi,
but the first time something  strongly attenuates the emission. 

This hypothesis finds support in the predictions of the two-peak accretion model. 
Taylor et~al. (1992) and Mart\'{\i} \& Paredes (1995)
have modeled the properties of \lsi in terms of an accretion rate
$\dot{M} \propto {\rho_{\rm wind}\over v_{\rm rel}^3}$ (where $\rho_{\rm
wind}$ is the density of the Be star wind and $v_{\rm rel}$ is the relative
speed between the accretor and the wind), which creates two peaks because of
the high
eccentricity ($e=0.7$).
The highest peak corresponds  to
 the periastron passage, 
because of the highest density, while the second peak occurs when the drop in the relative
velocity $v_{\rm rel}$ compensates (because of the inverse cube dependency)
the decrease in density.
Mart\'{\i} \& Paredes (1995) have shown
that both peaks are above the Eddington limit and therefore one expects that
matter is ejected twice within the 26.5 d interval 
(an updated version of the accretion rate curve, for periastron passage at $\Phi$=0.23, 
 is given in Fig. 2 of Bosch-Ramon et al. (2006)).

The two-peak accretion model has found  confirmation through the observations in another X-ray binary system, \object{Circinus X-1}, which also has a rather high eccentricity, e=0.8 \cite{murdin1980}. The observations by Tudose et al. (2006) show indeed a radio peak at periastron and a  second peak around apoastron.
The important difference between Circinus X-1 and \lsi is the nature of their respective companion stars. The companion star of Circinus X-1 is clearly cooler than a B5 star \cite{jonker2007} whereas in \lsi it is a B0 V star \cite{casares2005}.
In \lsi the ejected relativistic electrons
near periastron are embedded in such a strong UV-radiation field
that they nearly completely loose their energy through inverse-Compton scatterings.
This hypothesis found both theoretical and observational support:
Bosch-Ramon \& Paredes (2004) proposed a numerical model based on inverse Compton
scattering, where ejected  relativistic electrons  are exposed to stellar
photons (external Compton) as well as to synchrotron photons (synchrotron self
Compton) and  found indeed a gamma-ray peak at periastron.
The hypothesis of a high-energy outburst at periastron
 was  confirmed (Massi et al. 2005) by observations  
from the EGRET instrument (i.e., $E> 100$ MeV) on board of the Compton
 Gamma-Ray Observatory satellite (Tavani et al. 1998).
Massi et al. (2005) pointed out that the EGRET gamma-ray emission  during one full orbit (at JD 2450334, $\Theta$=0.18) shows
a clear peak at periastron passage and in a
previous epoch (at JD 2449045, $\Theta$=0.41) 
 shows an increase again near the periastron passage and  a second
   large peak at $\Phi\simeq 0.5$.
The inverse-Compton losses that produce the high-energy peak at periastron
seem to be  severe since
 in only very few radio light curves
 can one  distinguish a
peak at periastron. This is the case in
the radio observations that have been simultaneously performed with the MAGIC Cherenkov high-energy telescope. In Fig.~S1 of Albert et al. (2006),
one can see together with the radio peak at phase $\Phi = 0.7$ a second radio peak close to the periastron passage.

The  orbital phase of the two EGRET  peaks, $\Phi=$0.2 and $\Phi\simeq$0.5 agree with the  
two optically thick intervals
0.22$\pm$0.08 to 0.33$\pm$0.12  and from 0.50$\pm$0.09 to 0.70$\pm$0.13
determined here. 
The separation  between
the  centers  of the  two optically thick
intervals ($\Phi \simeq 0.3$   and   $\Phi= 0.6$)
agrees also with  the separation, i.e.  $\Delta \Phi=0.3$,
 between the two accretion peaks
of the model by Marti $\&$ Paredes (1995), who
adopted parameters for the envelope of the  Be star
derived from infrared data.
Romero et al. (2007) applied a smoothed particle hydrodynamics
 code to develop  three-dimensional, dynamical simulations for \lsi and found
that the  accretion rate has two peaks per orbit, i.e., a narrow peak at periastron ($\Phi=0.23$),
 and a broad peak that lags the periastron passage by about 0.3 in phase.
Finally, recent observations with VERITAS ($E >$ 500 Gev, Acciari et al. 2009) give
 marginal evidence for emission at $\Phi=$0.2-0.3 along with
the known  emission  at  $\Phi=$0.5-0.9.

Gamma-ray observations, the  two-peak model (fitted to infrared observations) and finally three-dimensional simulations
give  a separation between the two  peaks in  agreement with
the separation between the intervals we discussed, where 
the radio spectral index is optically thick.
The intervals
(0.22$\pm$0.08 to 0.33$\pm$0.12  and from 0.50$\pm$0.09 to 0.70$\pm$0.13)
determined here
from a database of 6.7~yr, therefore lead to important observational support,
at radio wavelengths,
for  the two-peak accretion/ejection model
 by Taylor et al. (1992) and 
Mart\'{\i} \& Paredes (1995).

\section{Spectral Index vs. Jet-States}

\label{classII}

A telling  definition of  
the two distinct radio emission states,  steady jet and transient jet, 
is  radio emission ``attached to" and ``detached from" the nucleus
(Dhawan et al. 2000).
In fact,  the steady jet (flat or inverted radio emission spectrum)
appears as a  compact jet centered on the engine, whereas the transient jet (emission with an optically thin spectrum) is associated with components   
moving with  relativistic speed away from the binary core
(Fender et al. 2004).

In Fig.~\ref{oo} $\Theta$=0.8-0.9,
 we indicate with two bars,
where VSOP observations were performed by Taylor et al. (2000, 2001).
The VSOP observations occurred  within the second
optically thick interval. 
The optically thin outburst,  $Peak_2$, connected
with the transient jet occurred 24 h after the end
of the VSOP observations.


The implication of this fact is that the set of VSOP images 
(like that shown  in our Fig. 1) used in the past to estimate 
the ``expansion" velocity,   show only a core centered, slow, steady jet.
 This explains the low velocity values derived.

The spectral index analysis therefore is a powerful tool to distinguish
between  steady and  transient jets. We can better  prove this  
by comparing  the spectral index analysis with  the different 
displacement of  radio structures imaged along the orbit of \lsp.
To compute such a displacement we use 
the astrometric results by Dhawan et al. (2006) for their radio features A,B,C,D,E,F,G,H,I and J.
We also use the orbit of the system from Casares et al. (2005)  
and  assume that optical and radio frame overlap each other  in B. 

\subsection{Spectral Index Analysis vs. Astrometry}
The astrometry of the radio peaks of VLBA observations 
at 3.6 cm is shown in 
Fig.~4 of Dhawan et al. (2006). In the figure
   the orbit is also given  as  
an ellipse of axis 0.5 AU  in an arbitrary  position.  
The exact location of
the ellipse relative to the radio measurements is, as mentioned by the authors,  in fact uncertain.

We   adopt for   the optical frame 
the orbit presented in Fig. 2 from Casares et al. (2005), and  
plot in Fig.~\ref{offset}-Top  
the distance of B (i.e. the orbital phase of B, shown as empty squares) 
in respect to all other orbital phases
corresponding  to the   VLBA  observations.
In the same figure we plot the relative displacement (as filled circles) of all the radio peaks 
reported in Fig. 4 of Dhawan et al. (2006) in respect to B, showing the offsets with an error bar of 
0.1 mas, as given by these authors. 
As a result,  the distance between each circle and
 its relative square gives an estimate of the 
displacement  of the radio structure  from the orbit.
The result of this comparison is 
that the  radio peaks C and J  are found  to be  on the orbit
whereas the radio peak A is  clearly offset (by  5$\sigma$) 
in the same way as E,F,G and H.
The offsets of the two radio peaks D and I,
displaced $< 3 \sigma$ away from  the expected orbital position,
are not significant for the analysis.
In Fig.\ref{offset}-Bottom we show $\alpha~vs~\Phi$ for the subset of the
 GBI database, 
 $\Delta \Theta$=  0.32-0.35,  that corresponds to  the  interval
of the Dhawan et al. (2006) observations, but at other epochs.
Considering that the data refer to a different epoch the agreement between the two plots is rather good.
The orbital phase interval, where we find emission centered on the orbit (i.e.
B,C and J),
is where  $\alpha \geq$ 0 (i.e. a steady jet).
Otherwise the emission clearly is optically thin for the rest of the orbit and 
this corresponds to the orbital phases of   E,F,G and H. 

\subsection{Relativistic Doppler Effects}
The radio peak A is offset (5$\sigma$) 
and the spectral index of Fig.\ref{offset}, even if not at the same epoch of the Dhawan et al (2006) observations, shows  indeed minor transients in the optically
thick interval.
This  scenario  finds  confirmation in  the Dhawan et al.  (2006)
image  C (see Fig.\ref{radiomaps}),  
where components, detached from the nucleus,  are observed at
the extremes of an already reformed steady jet. 

A central  instability driving  shock waves  into a bipolar outflow
should create  two symmetrical components of shocked material;
the lack of a receding component
for the radio structures  A, E,F,G and H indicates a large Doppler de-boosting factor. 
Doppler  effects are  noticeable in C (here Fig.~1): 
not only the reformed steady jet is clearly
 asymmetric, but consistently also
the distance of the ``plasmoids" from the core is different, with
$d_{a}$ for the approaching component  larger than $d_{r}$ for
the receding component.
Inserting the specific values that we get from 
Fig.~\ref{radiomaps}
into the following relation \cite{paredes2002}:
\beq
\beta \cos \theta={(d_{\rm a}-d_{\rm r})\over(d_{\rm a}+d_{\rm r})},
\eneq
we derive  $\beta \cos \theta \simeq$ 0.3.
This implies a minimum value  $\beta \simeq  0.3$ for small values
of $\theta$, the angle between the jet and the line of sight.

With respect to the morphology of the structures we note,
that even though some of them may still look like centered 
 one-sided steady jets, 
this is not in contradiction with the fact that they are 
far away from the center of the system, instead. 
This kind of  morphology was also found in the microquasar GRS 1915+105 and
can be seen in the last image of Fig.~2 from Fender et al. (1999),
which clearly shows two-sided  ejections (C2, C3, and  NW), where the 
offset structures  look like one-sided jets very similar to the radio 
structure  F at $\Phi= 0.600$ in Fig.~3 from Dhawan et al. (2006). 
That means the morphology  can be rather misleading.

\section{Spectral Index vs. X-ray States}

Steady jets and transient jets,
present in microquasars, are produced when the X-ray binary is in one of the  two different
 X-rays states:  low/hard and steep power-law state (Fender et al.  2004).
The  low/hard X-ray state  is the state characterized by a  
 power-law with photon index $1.5< \Gamma < 2.1$
 \cite{remillard-mcclintock2006} and
 a clear steep cut-off near 100 keV \cite{grove1998}.
The steep power law X-ray state
(formerly called very high state) was  renamed after monitoring
programs of ${\it RXTE}$  showed that, whereas an unbroken steep power-law
is a fundamental property of the state, a very high luminosity is not. 
The photon index in the X-ray and gamma-ray bands is the same: $\Gamma \sim 2.5-3.0$ 
\cite{mcclintock-remillard2006}. 

\subsection{Optically Thin Radio Spectrum and Steep Power-Law X-ray State}
Whereas there are  MAGIC observations where \lsi remains  undetected, it
is significant that when it is detected, 
 the energy spectrum is always well fitted by a power-law
with  photon index   equal to $\Gamma \simeq$2.6.
Albert et al. (2008a,b) discuss that although there is  clear evidence for a significant change
in the  flux level for two observations at two different epochs and at
$\Phi=$0.5$-$0.6  and  $\Phi=$0.6$-$0.7, 
the photon index during the first observation 
was  $\Gamma = 2.7 \pm 0.4$
and for the second observation was   $\Gamma = 2.6\pm0.2$.
This is consistent with the steep power law state, where the photon index is
a fundamental property of the state and not the luminosity \cite{mcclintock-remillard2006}.
Chernyakova et al.  (2006)  
analyzed INTEGRAL data from the time interval 2003--2005.
This interval corresponds to $\Theta$ from $\sim0.6$ to 
 $\sim 1.0$, i.e. during the maximum of the 1667~d cycle, where the
two optically thick intervals  cover the phases from 
 $\Phi$=0.2 $\pm$ 0.1  to  0.6 $\pm$ 0.1.
A steep power-law X-ray state is expected
where  $\alpha$ declines under zero, i.e.  
after $\Phi=0.6 \pm$0.1.
Indeed, in the orbital phase interval $\Phi=0.6-0.8$,
Chernyakova et al.  (2006) determine 
 $\Gamma=3.6^{+1.6}_{-1.1}$. 
 If $\Phi=0.4-0.6$ they determine
 $\Gamma=1.7\pm0.4$ and if $\Phi=0.8-0.4$  $\Gamma=1.4\pm0.3$.
EGRET observations at $\Theta$=0.18,
shows 
the peak predicted by the two-peak accretion model
 at periastron passage.
However, EGRET data 
at  $\Theta$=0.4, like MAGIC observations, show a second  
broad  peak at $\Phi\simeq 0.5$, as would be expected 
between X-ray and very high energy,  in the case of an unbroken power law. 
MAGIC  observations simultaneous with INTEGRAL observations are not available
and therefore
fluxes cannot be extrapolated and compared. However, it is a fact that
the measured photon indices in the X-ray and gamma-ray bands
overlap the expected range for the steep power-law state and that
this occurs in the orbital range
where, following our spectral index a  steep power-law state
is expected.

Simultaneous observations of MAGIC 
 and INTEGRAL exist for
Cygnus X-1 (Malzac et al. 2008 and reference there). The   photon index of MAGIC observations fitted
by a power-law is $\Gamma=3.2 \pm 0.6$.  INTEGRAL observations
have only be fitted with a power-law model plus cut-off (i.e. as a low-hard)
and therefore cannot be compared. However, the  very important
result is that TeV emission can be extrapolated
down to the MeV range assuming $\Gamma= 2.5$ (Fig. 6 in Malzac et al. 2008).
The transition in Cygnus X-1  from
its low X-ray state and ``high X-ray state" is a transition
from low-hard state with a breaking type spectrum with $\Gamma\sim$1.8,
and a  steep power-law state with  $\Gamma=2.6$ (Figs. 2 and 3 in
McConnel et al. 2001).
The steep power-law state is ``weird"  with its   low luminosity
and the absence of  QPO 
(Mc Clintock and Remillard 2006 and references there).

A steep power-law state could be 
the origin of the TeV-emission for
both \lsi and Cygnus X-1.
The physical process could be  synchrotron self Compton   scattering 
by the shock-generated  population
of relativistic electrons responsible for the radio emission.
In Fig.~3 the spectral index at the peak of the  optically thin
      outburst, $Peak_2$, is $\alpha\simeq$-0.5.
Bosch-Ramon \& Paredes (2004) calculated that an electron
 power-law  index $p=1.7$, and therefore $\alpha=-(p-1)/2=-0.35$, 
agrees  with a photon index $\Gamma$=2.2.
The steeper power law $p=2$, resulting from  our $\alpha \simeq$ -0.5,
implies   a steeper photon index, i.e. $\Gamma  >$2.2,
in  agreement with the values  
of the most sensitive observations performed
by MAGIC and VERITAS (2.6$\pm$0.2 and 2.4$\pm$0.2).

\subsection{Inverted  Radio Spectrum and Quiescent  X-ray State}

The X-ray luminosity $L_{\rm X}$ in the low/hard X-ray state
corresponds to a radiatively inefficient, ``jet-dominated'' accretion
mode (Fender et al. 2003).
In this mode only a negligible fraction of the binding
energy of the accreting gas is directly converted  into radiation
and   most of the accretion power emerges in kinetic
form, as shown for Cygnus X-1  by Gallo et al. (2005),
and for AGNs through  the relationship between the  Bondi power and the  
kinetic luminosity (Merloni \& Heinz 2007 and references therein).
For a compact object of a few solar masses the low/hard state
corresponds to $L_{\rm X} \sim 10^{36}~ {\rm erg/sec}$
but may drop to 
$L_{\rm X}= 10^{30.5} - 10^{33.5}$ erg/s \cite{mcclintock-remillard2006} 
at its  lowest  phase,   called 
quiescent state.  
The quiescent  state has several  characteristics of the low/hard state including
the presence of a continuous radio jet.  An example of such a situation is the radio
jet present in the black hole A0620-00 during its quiescent state \cite{gallo2006}.
There is a  relationship  between
the X-ray and radio luminosities of low/hard state black hole
X-ray binaries, that is  valid up to X-ray luminosities of
$10^{-8.5} L_{\rm Edd}$ (Gallo et al. 2003, 2006). This important non-linear
scaling between X-ray and radio emission has been demonstrated 
to hold - with the addition of a mass term - 
across the entire black hole mass spectrum, from microquasars to AGN  \cite{merloni2003,falcke2004}.
The X-ray luminosity of \lsp, $L_{\rm X}\simeq (1-6) \times 10^{34}{\rm erg/sec}$ \cite{paredes1997},  
is  between the quiescent and the low/hard states.
We checked if such a low luminosity, $L_{\rm X}\simeq 3\times 10^{34}{\rm erg/sec}$, correlates
with the  radio emission $L_{\rm R}=10^{31}$erg/s  \cite{combi2004},
measured outside the large radio outburst window.
The universal relationship for an object of about three solar masses predicts (Merloni et al. 2003): 
\beq
{\rm log}\ L_{\rm R}= \eta~ {\rm log}\ L_{\rm X} + 7.7
\eneq
with $\eta$ in the range from 0.5 to 0.7. 
Substituting  the observed luminosities of \lsi in radio and X-ray, $L_{\rm R}$ and $L_{\rm X}$, 
we obtain 
a value of $\eta$  consistent with the given  range, i.e. $\eta$=0.67.
Assuming an average value of  $\eta$=0.60, given by Merloni et al. (2203), it would result a radio luminosity of
almost three  orders of magnitude lower than the observed $L_{\rm R}=10^{31}$ erg/s. In other words, \lsi is 
more radio loud than ``average" microquasars (i.e. those corresponding to the center
value of  $\eta= 0.60$).
In the same way 3C 273 with $\eta \simeq $0.65 is more radio loud than
AGNs with  $\eta=0.60$ (see  Fig. 5 by Merloni et al. 2003).
\lsi also is extreme concerning other characteristics.
Several X-ray binaries show  quasi periodic oscillations with timescales of a fraction of minutes during the  low/hard X-ray state 
\cite{brocksopp2001}.
\lsi shares the remarkable property of extremely  slow quasi-periodic oscillations with the  black holes V404 Cyg and \grs.
During the decay of the 1989 outburst
a variety of   22$-$120 min  oscillations were observed in 
V404 Cyg (radio: Han \& Hjellming 1992).
Quasi periodic oscillations of 30$-$84 min  were
observed in \lsi  (radio: Peracaula et al. 1997  and 
 X-ray: Harrison et al. 2000).  
Radio observations of GRS 1915+108 by Pooley and Fender (1997)  revelead quasi-periodic oscillations with
periods in the range 20 $-$ 40 min and observations by Rodriguez and Mirabel (1997) oscillations
of 30 min.

Besides the low X-ray luminosity  the quiescent state has 
an  important characteristic that  is perfectly matched by \lsp.
This  important characteristic of the quiescent state
is its broadband spectrum with two
components: one in the X-ray band (represented by the  power law
with a photon index of $1.5< \Gamma < 2.1$
and a cut-off near 100 keV discussed above)
and the other is in the optical band (see the results from the  Hubble Space Telescope and Chandra
in McClintock et al. 2003).
 The observed optical/UV band emission comes
 predominantly from synchrotron radiation of relativistic
electrons present either in the ADAF\footnote{ADAF: Advection-Dominated Accretion Flow disk model} \cite{narayan1997, mcclintock2003} or in the jet \cite{russell2007}. In fact,  the steady jet is known to radiate at radio/infrared/optical wavelengths
\cite{russell2006, russell2007}. 
In this context the  optical observations of
LS~I~+61$^{\circ}$303
strongly support the hypothesis of a  quiescent state. 
Mendelson \& Mazeh  (1989, 1994)  discovered a clear period ($P=26.4$~d)
in optical observations. The authors noticed that the  
optical modulation has the same period as the radio outbursts and rises 
also in the same orbital phase reaching an amplitude of 
0.05 mag.
The authors suggested that the minimum is associated with the Be star, whereas the excess is created
by a periodic mechanism related with the accretion disk or with optical synchrotron radiation.
The photometry in the V band by Paredes et al. (1994) confirmed 
the periodicity and the shape of the optical light curve.
They found  a broad brightness 
maximum near the major radio outburst, i.e., $\Phi= 0.5-0.9$.
A new confirmation of the optical modulation  comes from the 
series of observations in the V and U bands spanning 6400 d by 
Zaitseva \& Borisov (2003). They found the optical period exactly equal to the radio one 
with a maximum at $\Phi=0.85$
and they noticed that the light curve has a gentler rise and a steeper decline.
The correlation between the optical and
radio emission of \lsi therefore exists in both 
period and orbital phase  and indicates the same agent: the  jet.

\section{Conclusions}
\label{conclusions}

The radio emission from \lsi
with  periodic ($P_1$=26.5~d)
 outbursts with an amplitude  modulated over $P_2=$1667~d
is detectable during the whole orbit.
If \lsi is a microquasar, the radio emission should show the characteristics
of such objects.
The current observational picture of jets distinguishes two completly
different types of radio emission, one  corresponding to a flat or inverted spectrum
and the other to an optically thin spectrum (Fender et al. 2004).
These two kinds of radio emission and their corresponding types of ejection
are related to each other.
The optically thin spectrum  is created by shocks caused by
 highly-relativistic plasma (transient jet) 
traveling  through a slower steady-flow  (steady jet) established 
during a previous phase of emission with an inverted spectrum (Fender et al. 2004). 
If spatially resolved, the flat/inverted spectrum 
 radio emission (steady jet)
appears as a continuous  jet centered on the system,
whereas the optically thin spectrum associated to outbursts (transient jets),
is resolved into components  (``knots" or ``plasmoids") moving with relativistic speed away from the binary core. 
These  two classes of  ejection correspond with two different X-ray states.
The transient jet is associated with the steep power law X-ray state, i.e. 
an unbroken steep power-law,
where the photon index in the X-ray and gamma-ray bands is the same and equal to 
 $\Gamma \sim 2.5-3.0$.
A steady jet corresponds always to 
the  low/hard  X-ray state 
 characterized by a
power-law with photon index $1.5< \Gamma < 2.1$
\cite{remillard-mcclintock2006} and
a clear steep cut-off near 100 keV \cite{grove1998}.

\begin{figure*}[t!]
\begin{center}
\includegraphics[scale=0.45, angle=-90.]{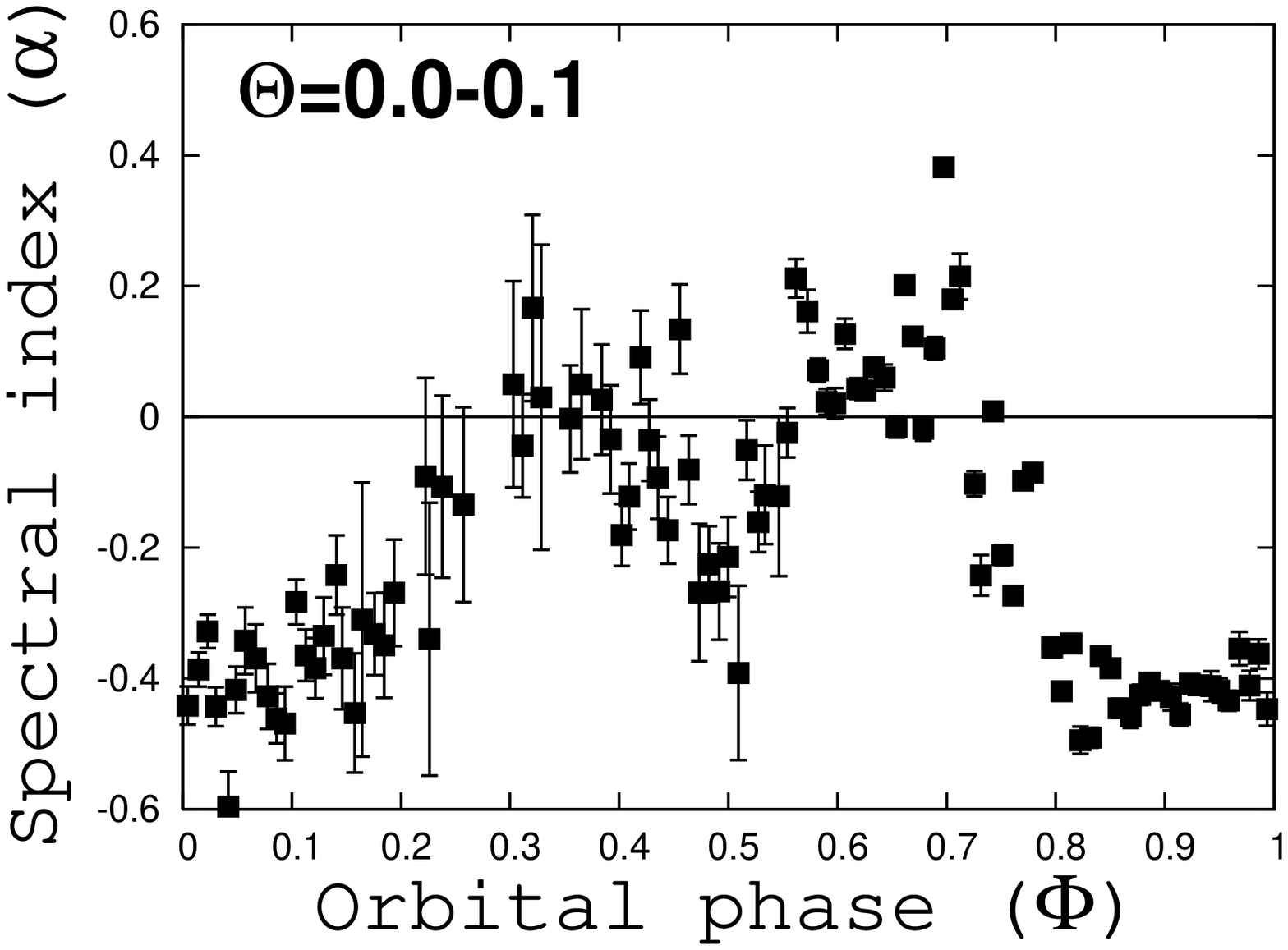}\\
\includegraphics[scale=0.45, angle=-90.]{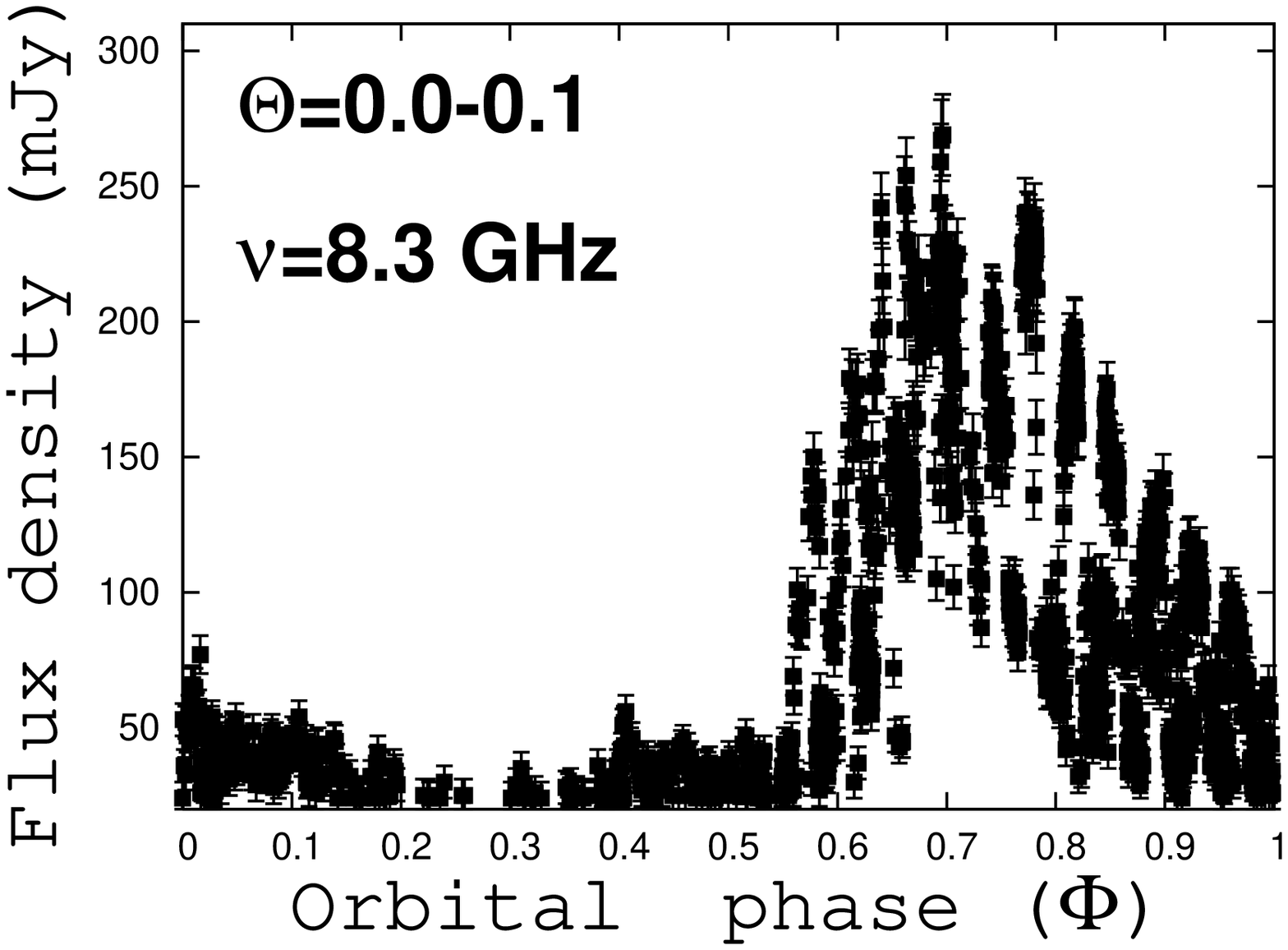}\\
\includegraphics[scale=0.45, angle=-90.]{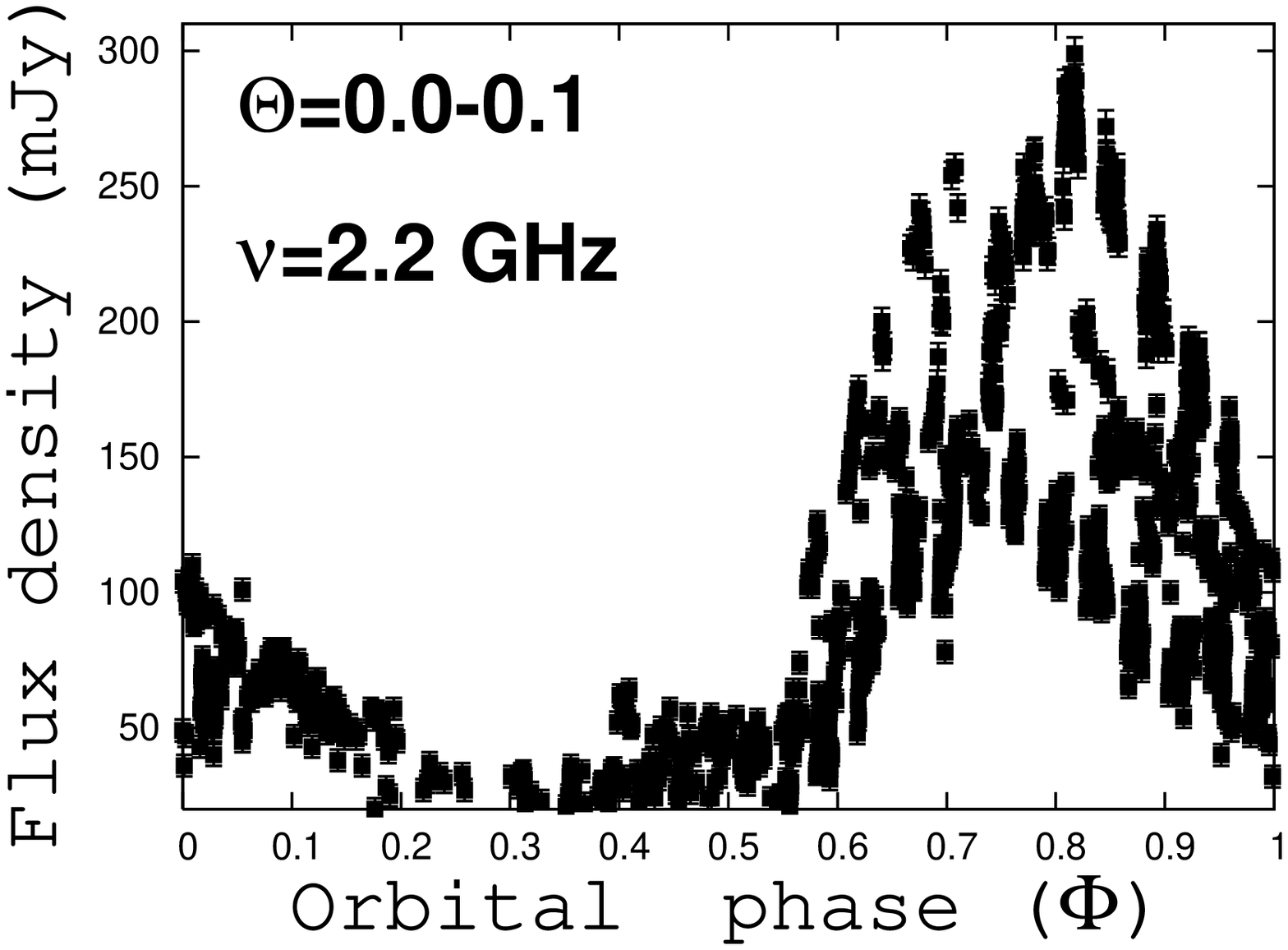}\\
\end{center}
\caption{
Spectral index and flux density, at 8.3 GHz and 2.2 GHz  vs orbital phase, $\Phi$,
for the GBI data in the interval $\Theta$=0.0-0.1. 
The spectral index,  $\alpha$, is  averaged over $\Delta \Phi$=0.009,  ($\sim$ 6 h).
Note that the evolution  from an optically thick to an optically thin spectrum
 occurs twice, giving
 the  $\alpha$ vs $\Theta$ curve  a double-peaked  shape.
The   radio outburst peaks first at 8.3 GHz, then it follows a 
larger  peak at 2.2 GHz.
Between the two outbursts there is
an   inversion of the spectrum from inverted to optically thin.}
\label{oo4}
\end{figure*}

\begin{figure*}[t!]
\begin{center}
\includegraphics[scale=0.4 , angle=-90.]{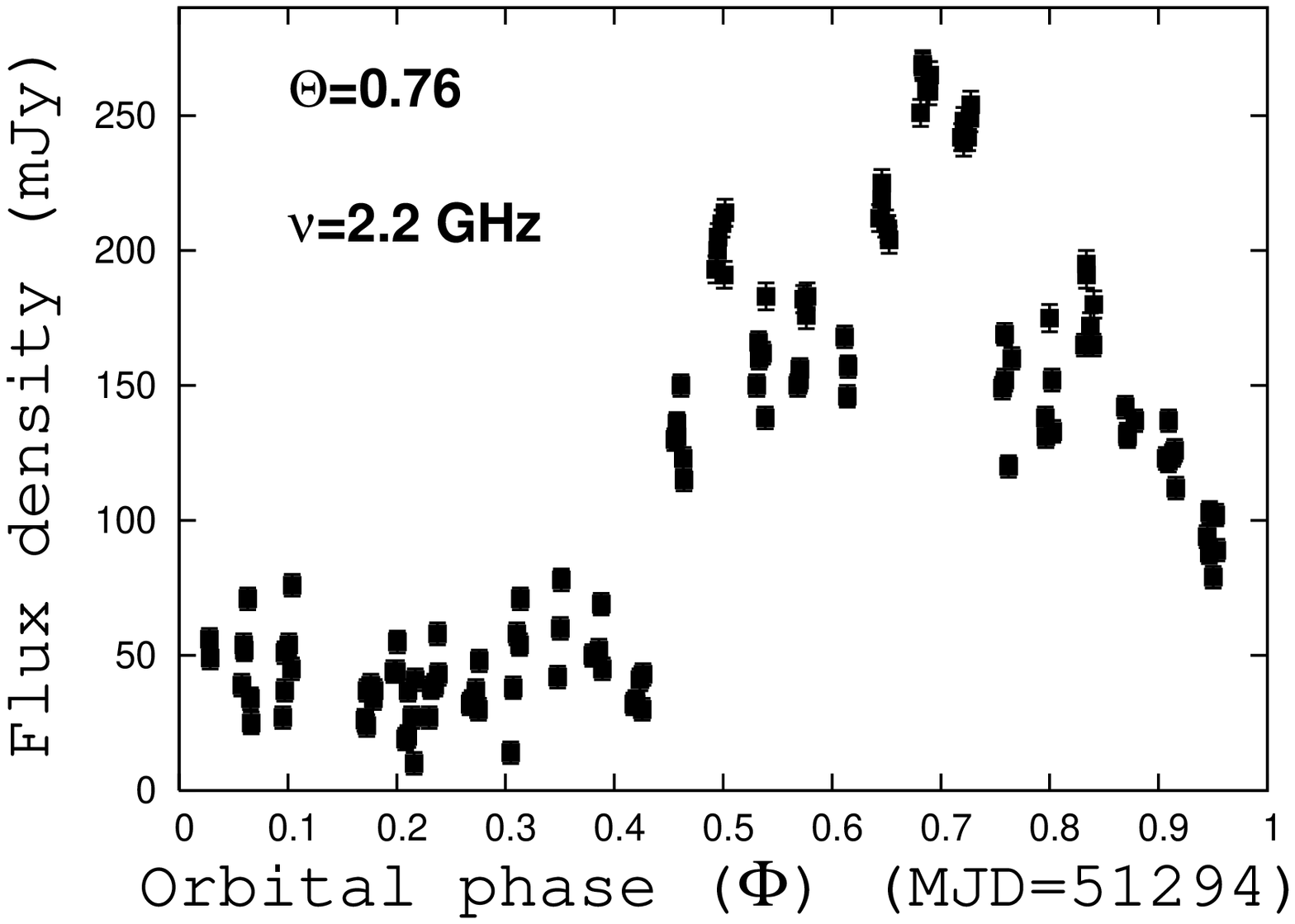}\hspace{0.5cm}
\includegraphics[scale=0.4, angle=-90.]{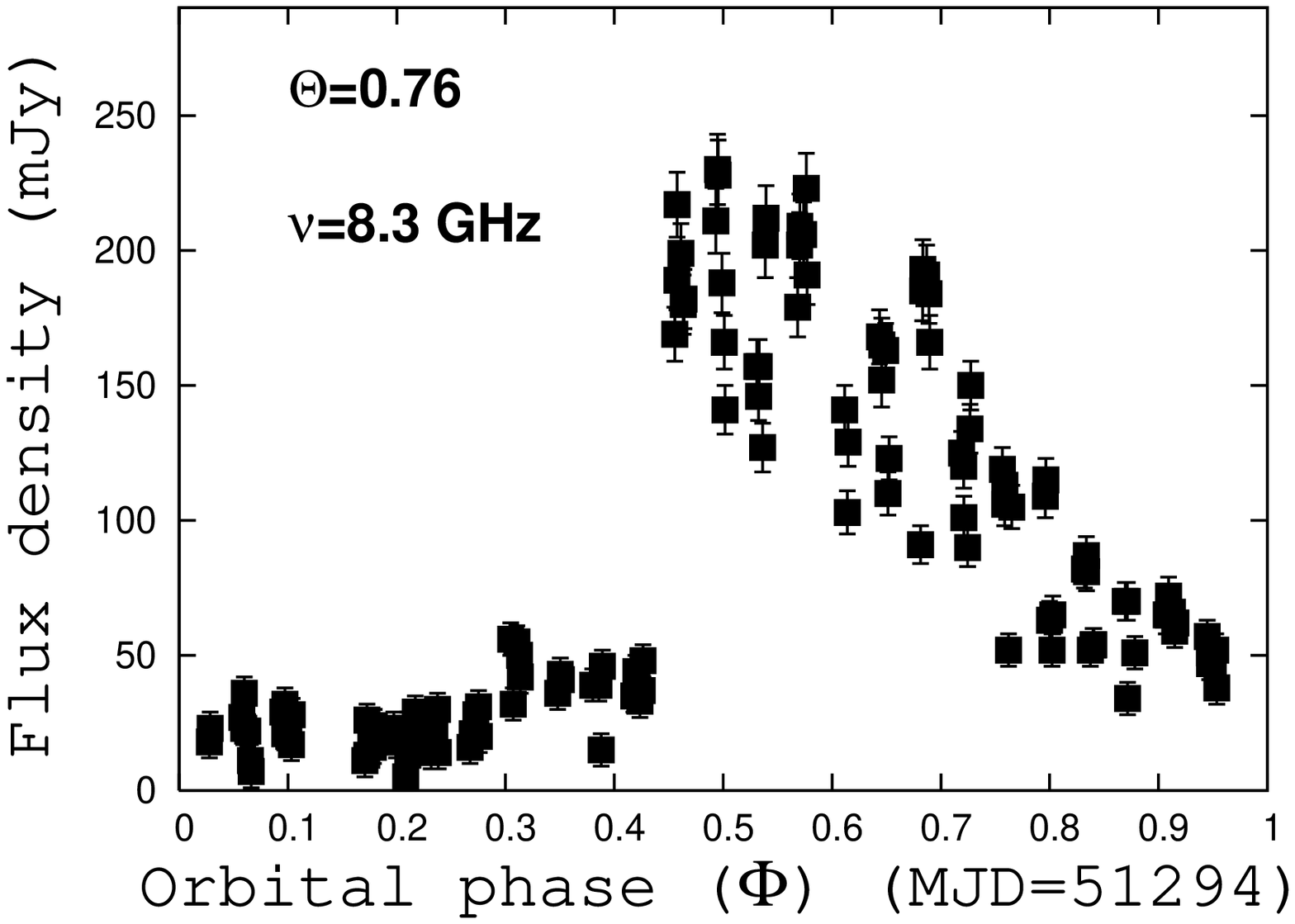}\vspace{0.5cm}
\includegraphics[scale=0.4, angle=-90.]{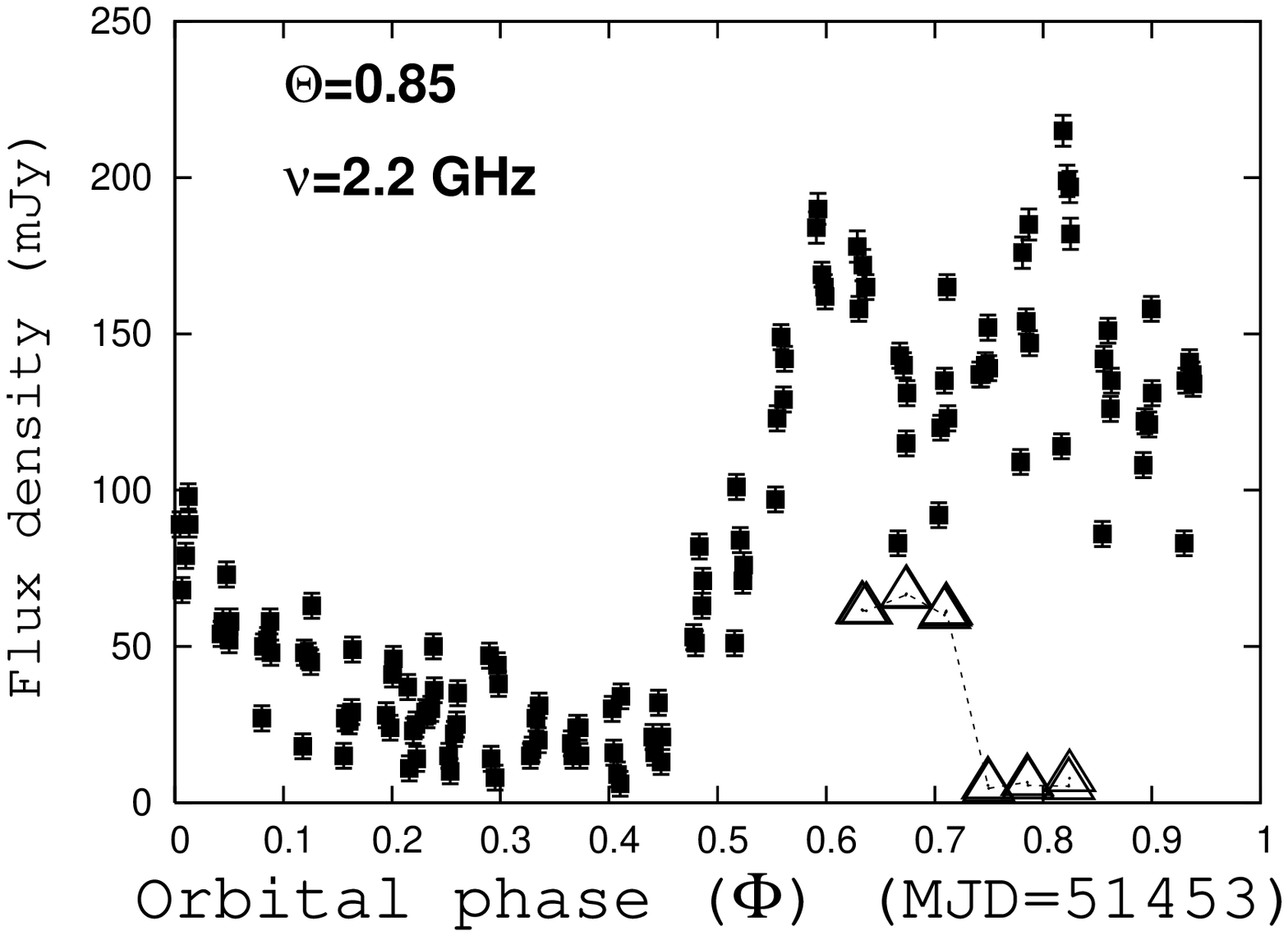}\hspace{0.5cm}
\includegraphics[scale=0.4, angle=-90.]{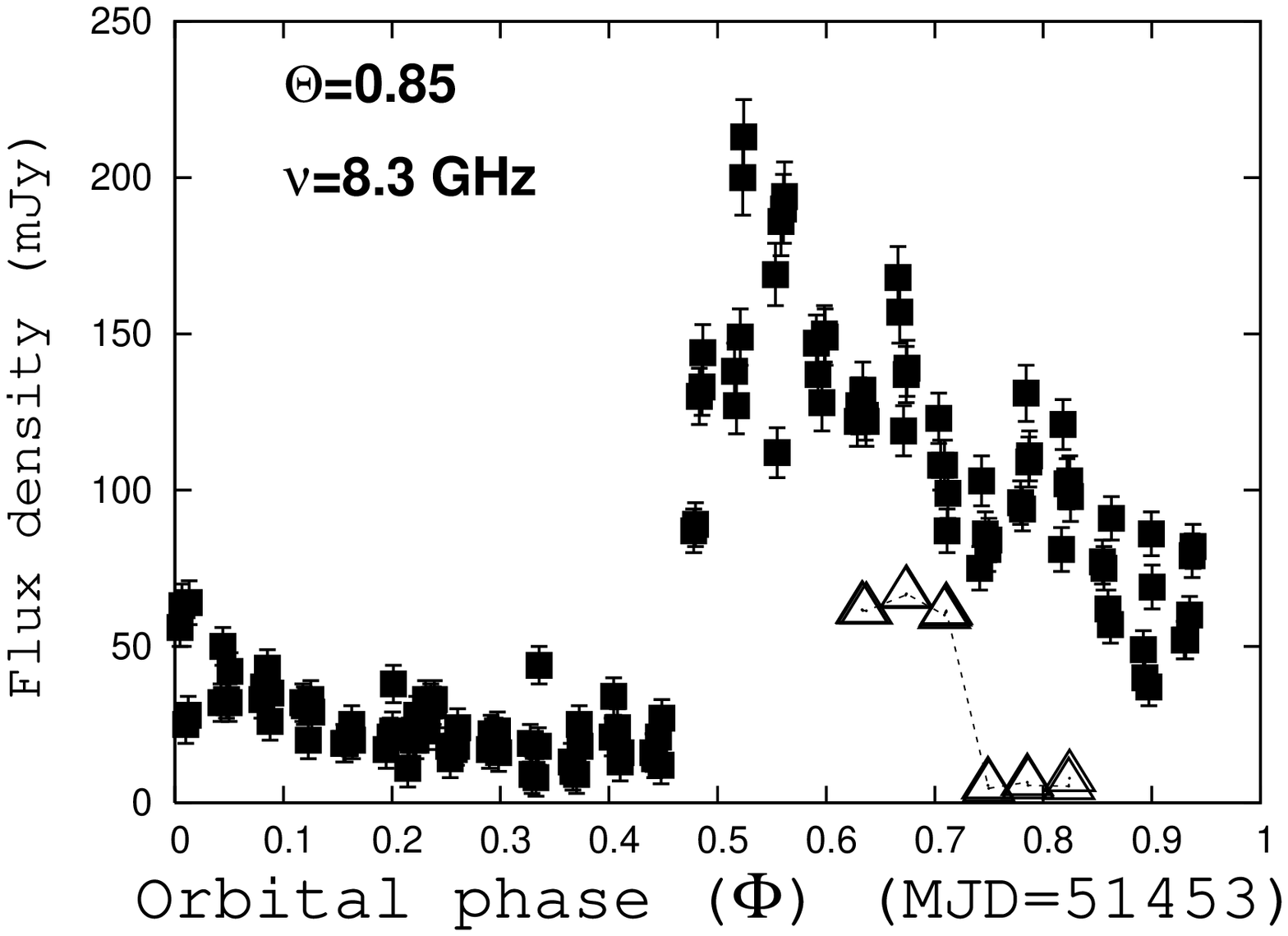}\vspace{0.5cm}
\end{center}
\caption{Top: Light curve at 2.2 GHz and 8.3 GHz at $\Theta$=0.76. 
Both light curves at 2.2 GHz and 8.3 GHz have two peaks: there are
two consecutive outbursts, one with optically thick  spectrum and the
second with optically thin spectrum. 
The  relative spectral index is shown in Fig.~5, panel 7.
Bottom: Light curves at  $\Theta$=0.85. 
The  relative spectral index is shown in Fig.~5, panel 8.
The triangles are the   H$\alpha$ emission-line 
measurements by Grundstrom et al. (2007, Fig.~3-Bottom) 
multiplied by a factor of -5 to fit in the plot. 
The H$\alpha$ value, still high at $\Phi=0.71$, 
registers a dramatic decline at $\Phi=0.749$ (1 day later)
in correspondence with the onset of the optically thin outburst.}
\label{oo5}
\end{figure*}

\begin{figure*}[t!]
\begin{center}
\includegraphics[width=.17\textheight , angle=-90.]{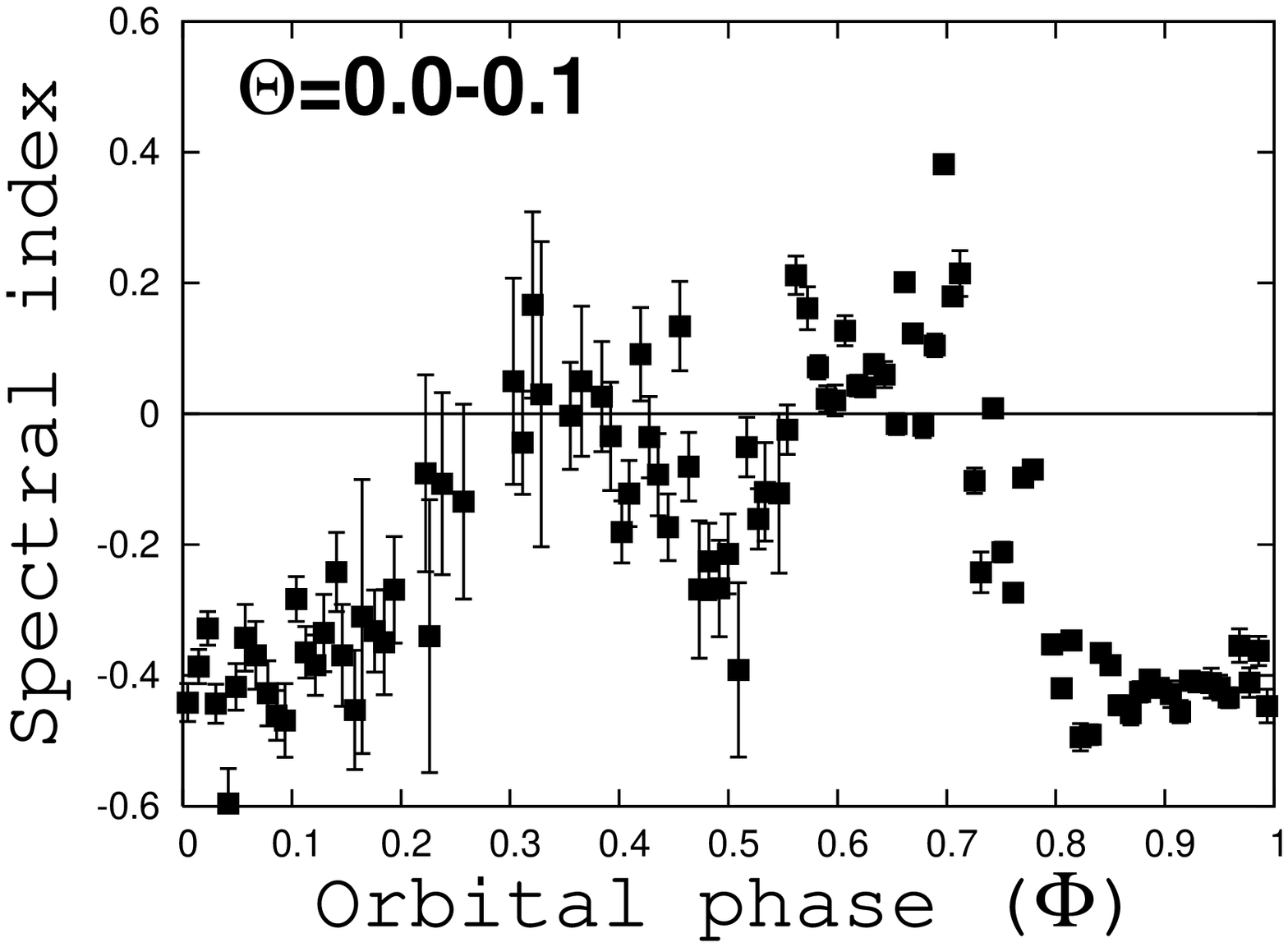}
\includegraphics[width=.17\textheight , angle=-90.]{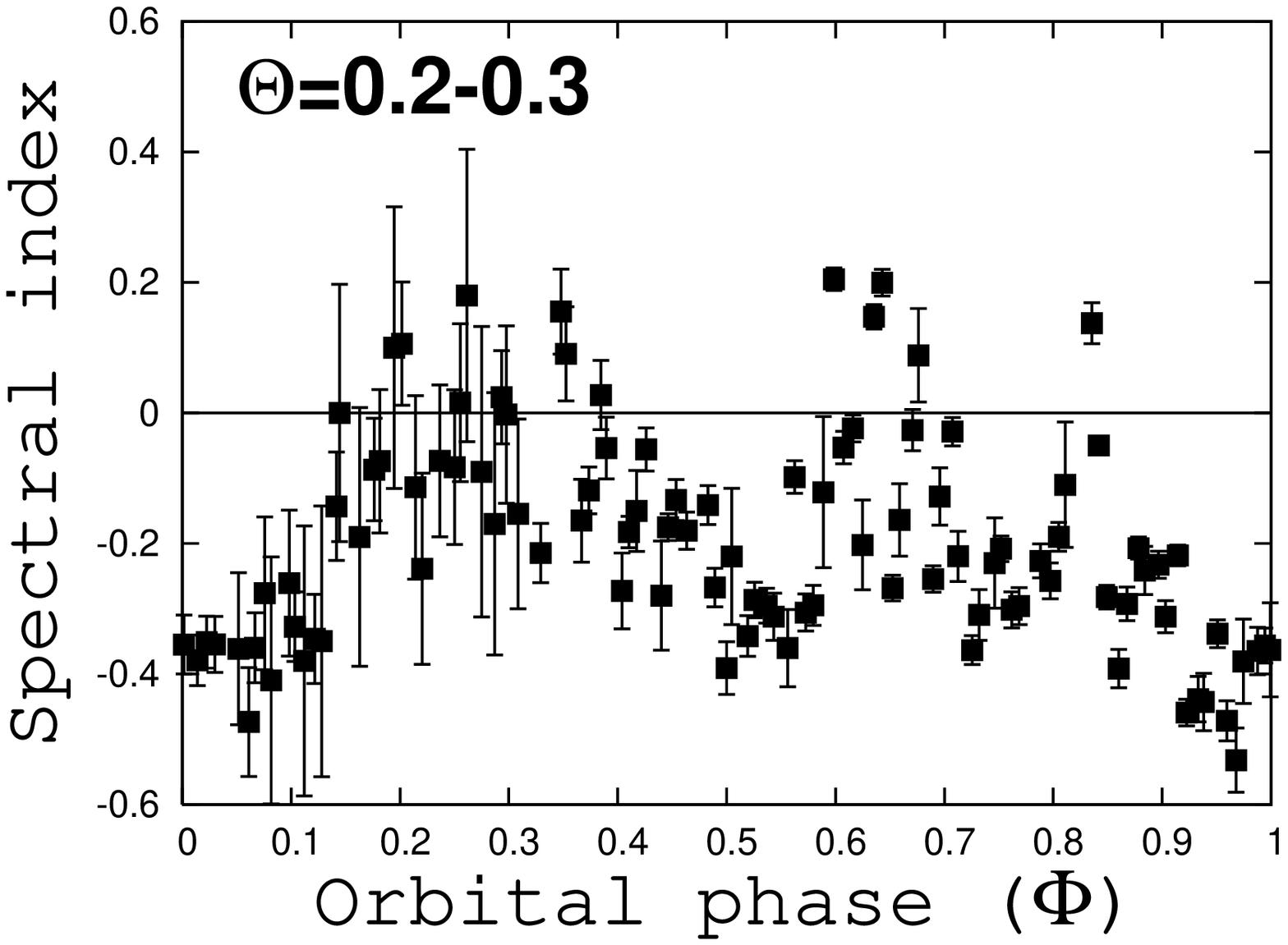}
\includegraphics[width=.17\textheight , angle=-90.]{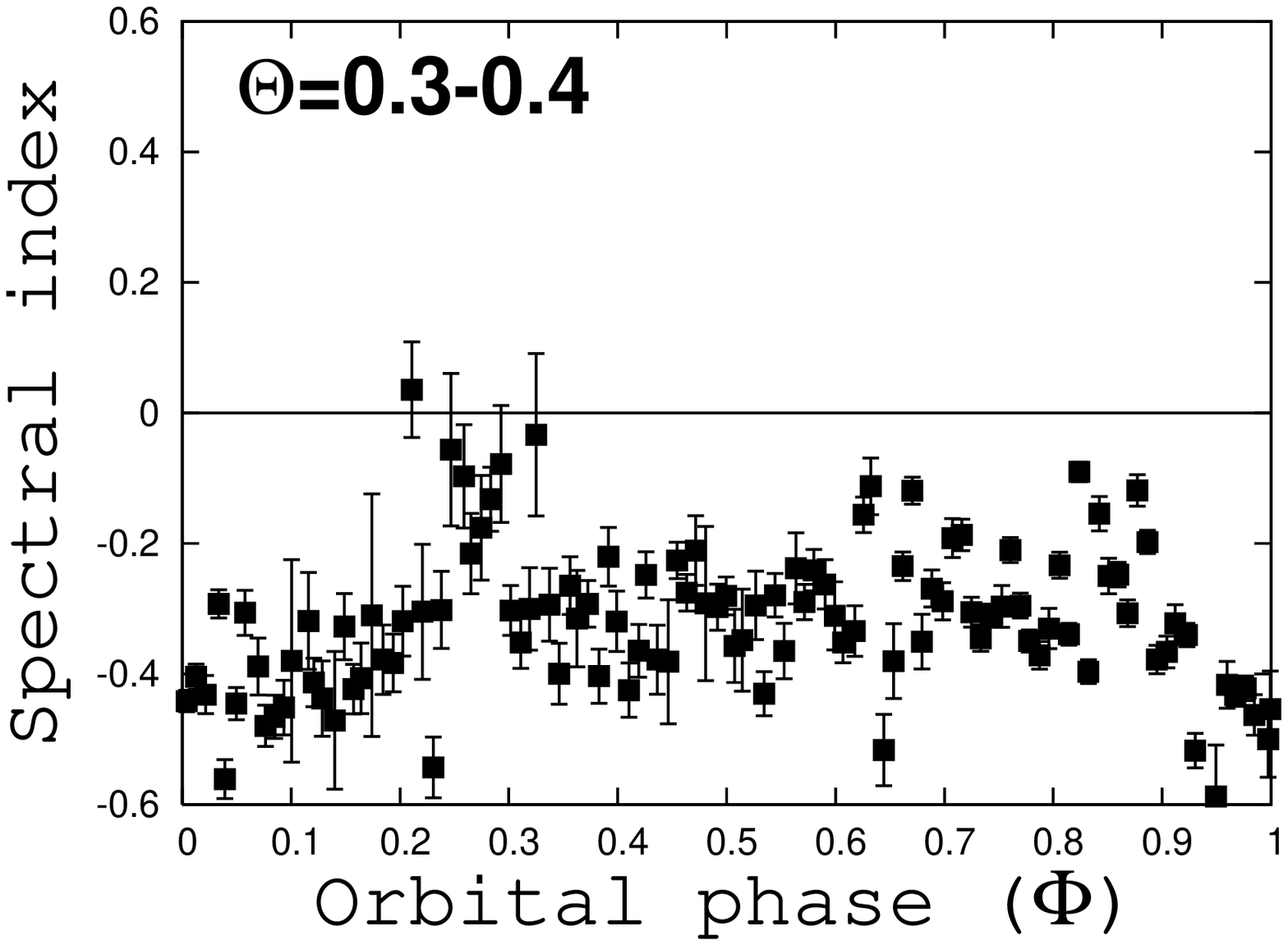}
\includegraphics[width=.17\textheight, angle=-90.]{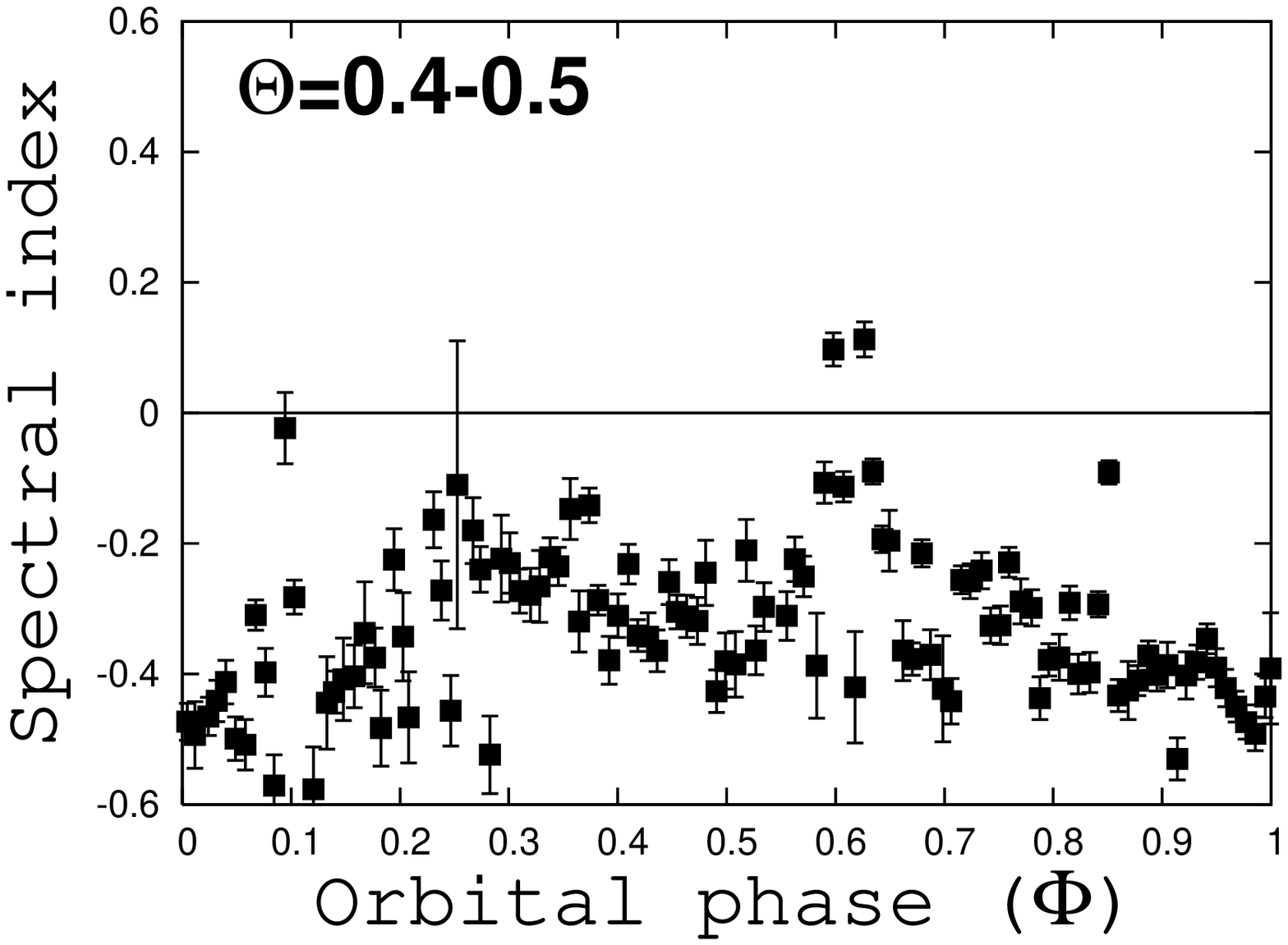}
\includegraphics[width=.17\textheight , angle=-90.]{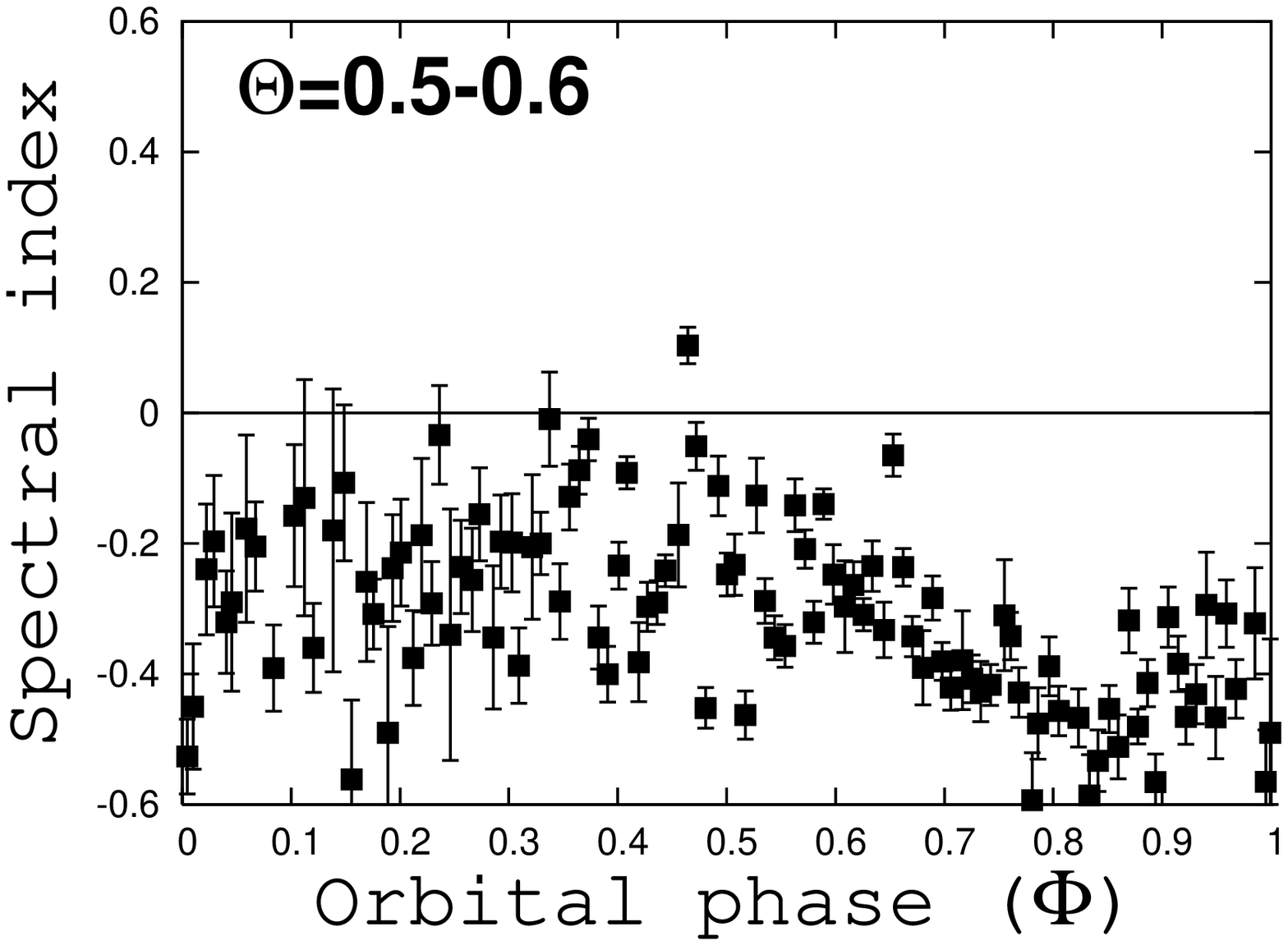}
\includegraphics[width=.17\textheight , angle=-90.]{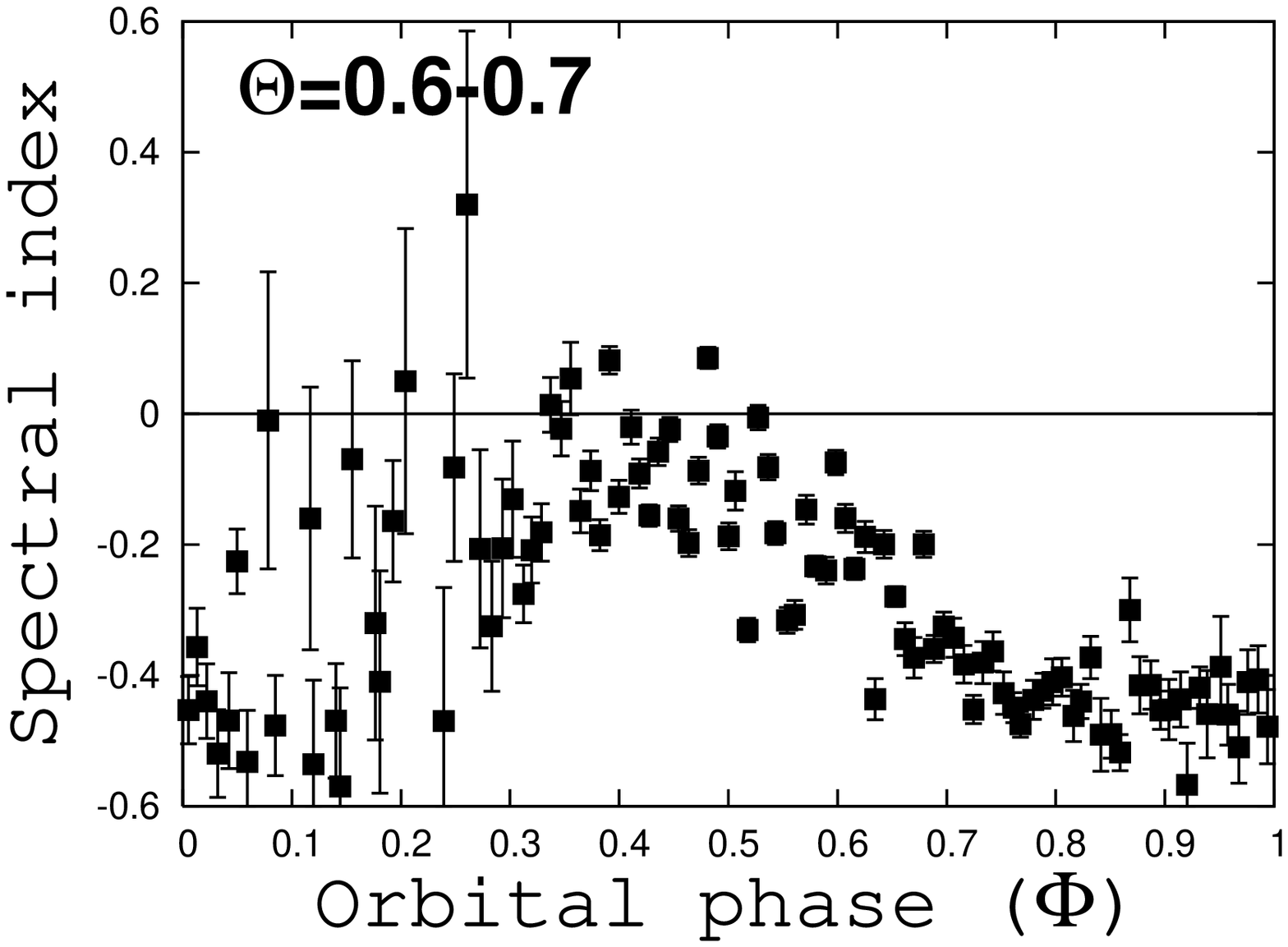}
\includegraphics[width=.17\textheight , angle=-90.]{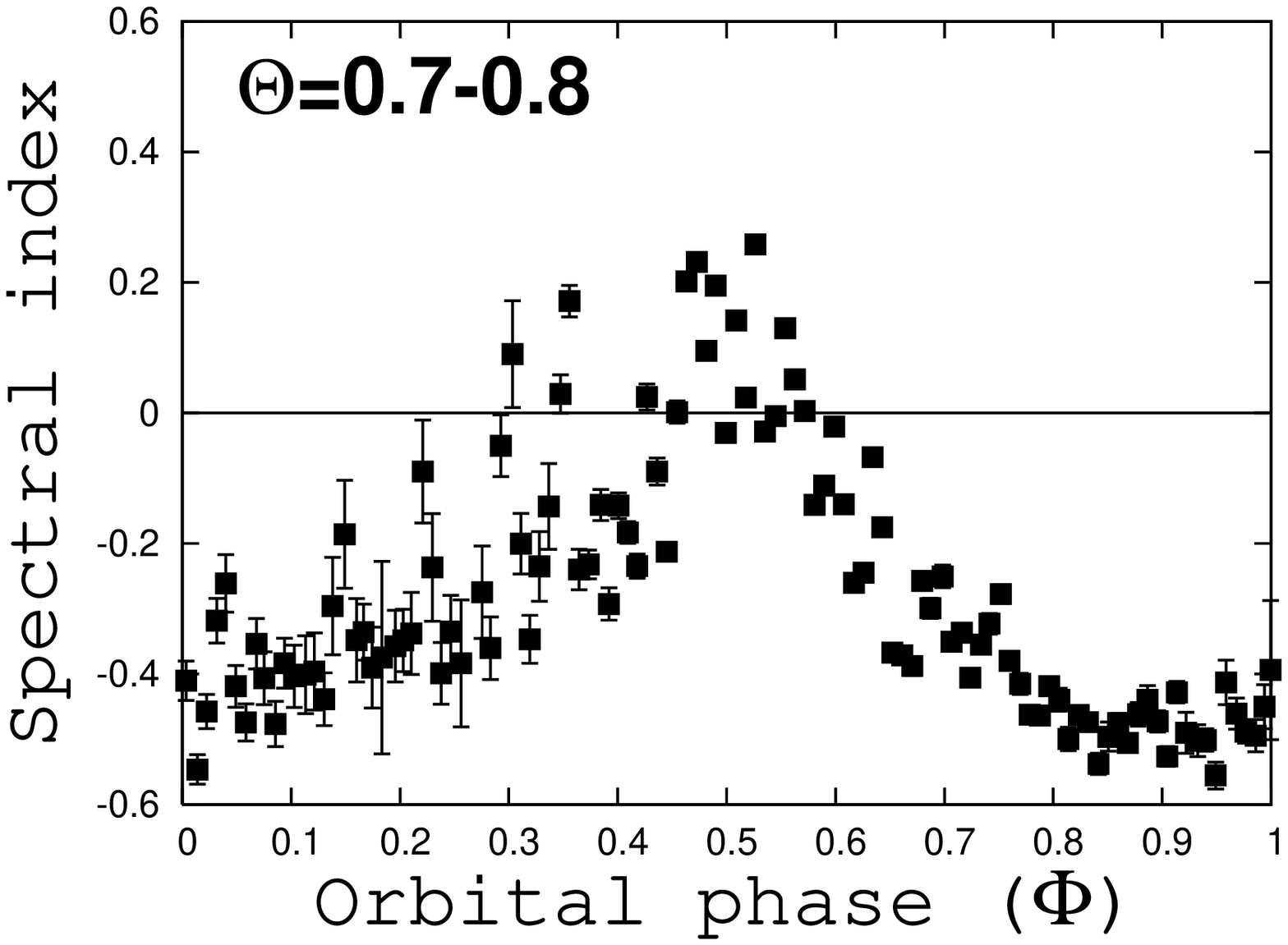}
\includegraphics[width=.17\textheight , angle=-90.]{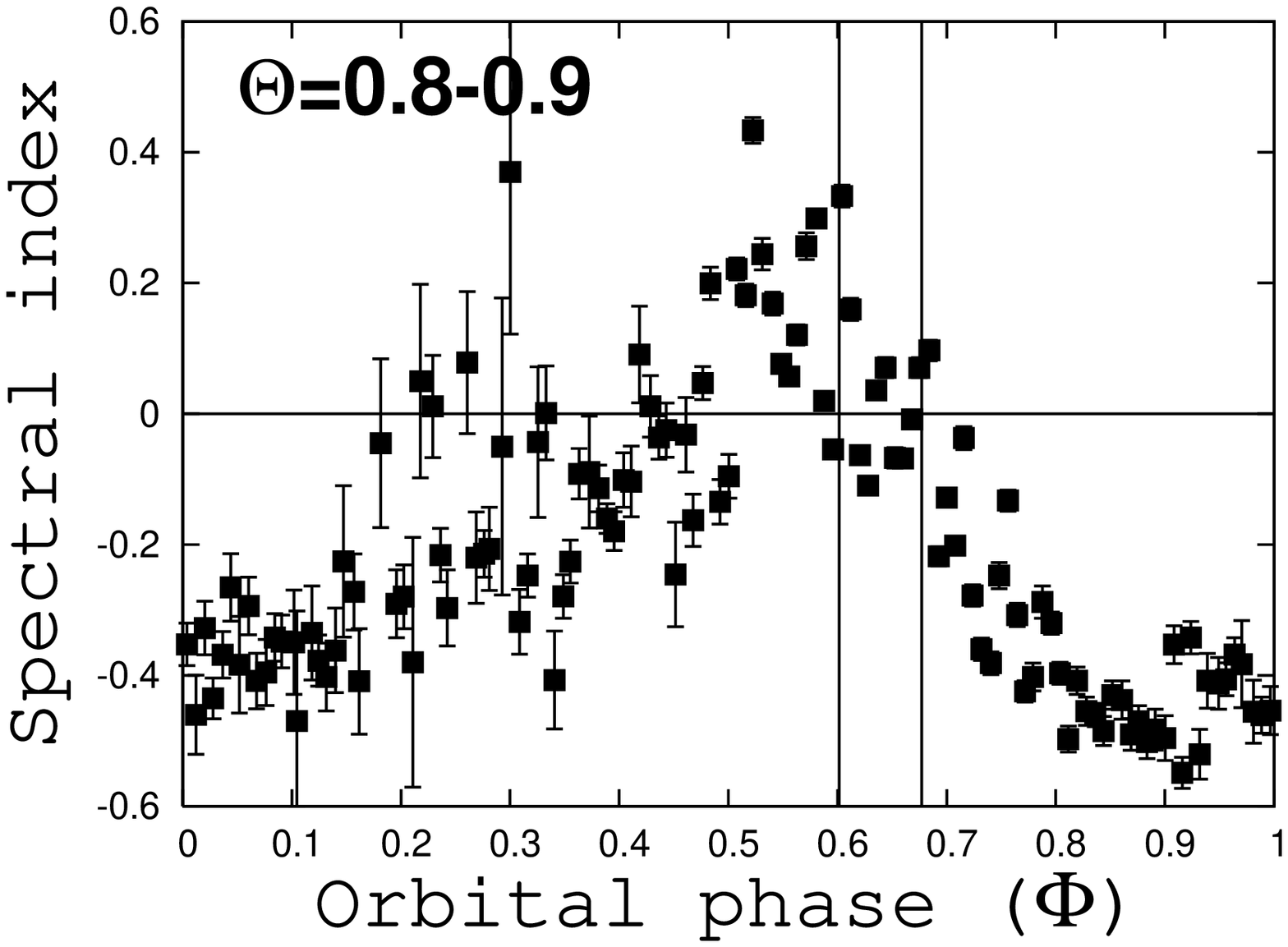}
\includegraphics[width=.17\textheight , angle=-90.]{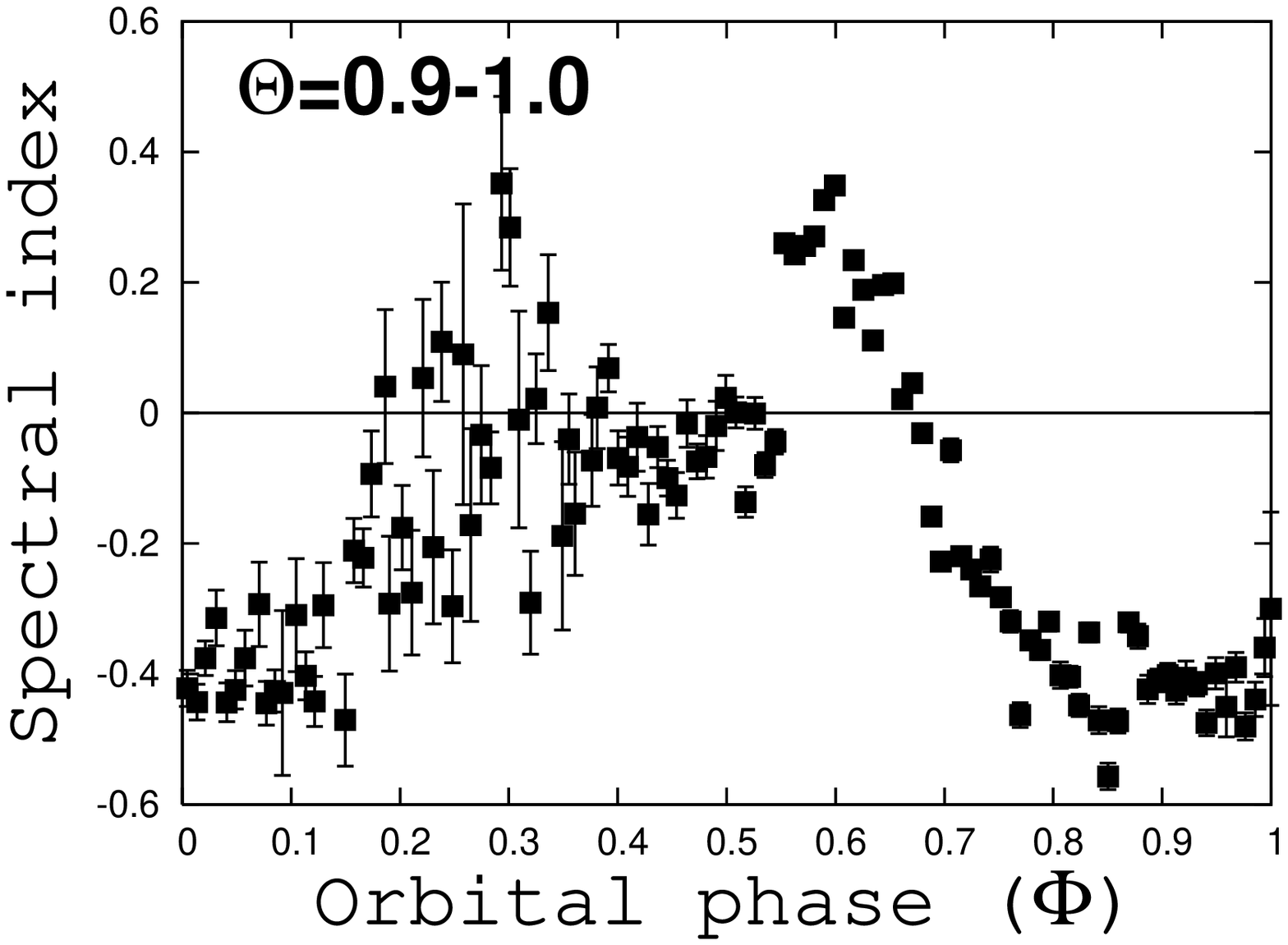}
\end{center}
\caption{
Spectral index vs orbital phase, i.e. $\alpha$ vs  $\Phi$, for different intervals of $\Theta$.
The spectral index is averaged over $\Delta \Phi$=0.009 ($\sim$ 6 h).
During $\Theta$=0.0-0.3 and  $\Theta$=0.7-1.0, i.e. during the maximum of the 1667~d cycle,
two optically thick intervals are present, i.e. intervals along the orbit where 
$\alpha \geq$0.  
The two optically thick intervals are from  
$\Phi$=0.22$\pm$0.08 to  $\Phi$=0.33$\pm$0.12 for the first one and, 
from $\Phi$=0.50$\pm$0.09 to  $\Phi$=0.70$\pm$0.13 for the second one.
The interval $\Theta$= 0.8-0.9  contains
two vertical lines showing the period when  the  GBI data were
simultaneous  with the
VSOP observations  by Taylor et al. (2001).
}
\label{oo}
\end{figure*}

\begin{figure*}[t!]
\begin{center}
\includegraphics[width=.17\textheight , angle=-90.]{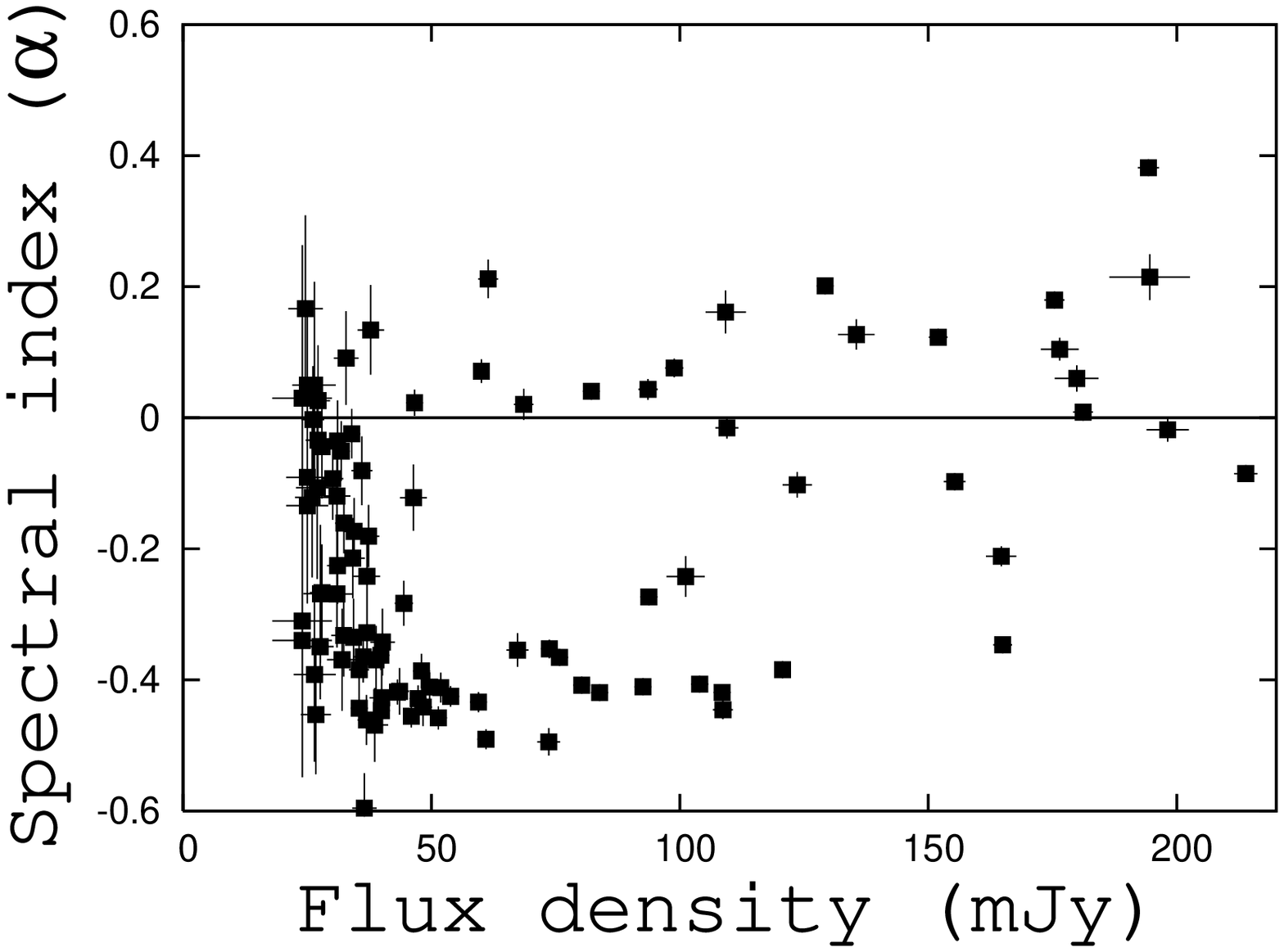}
\includegraphics[width=.17\textheight , angle=-90.]{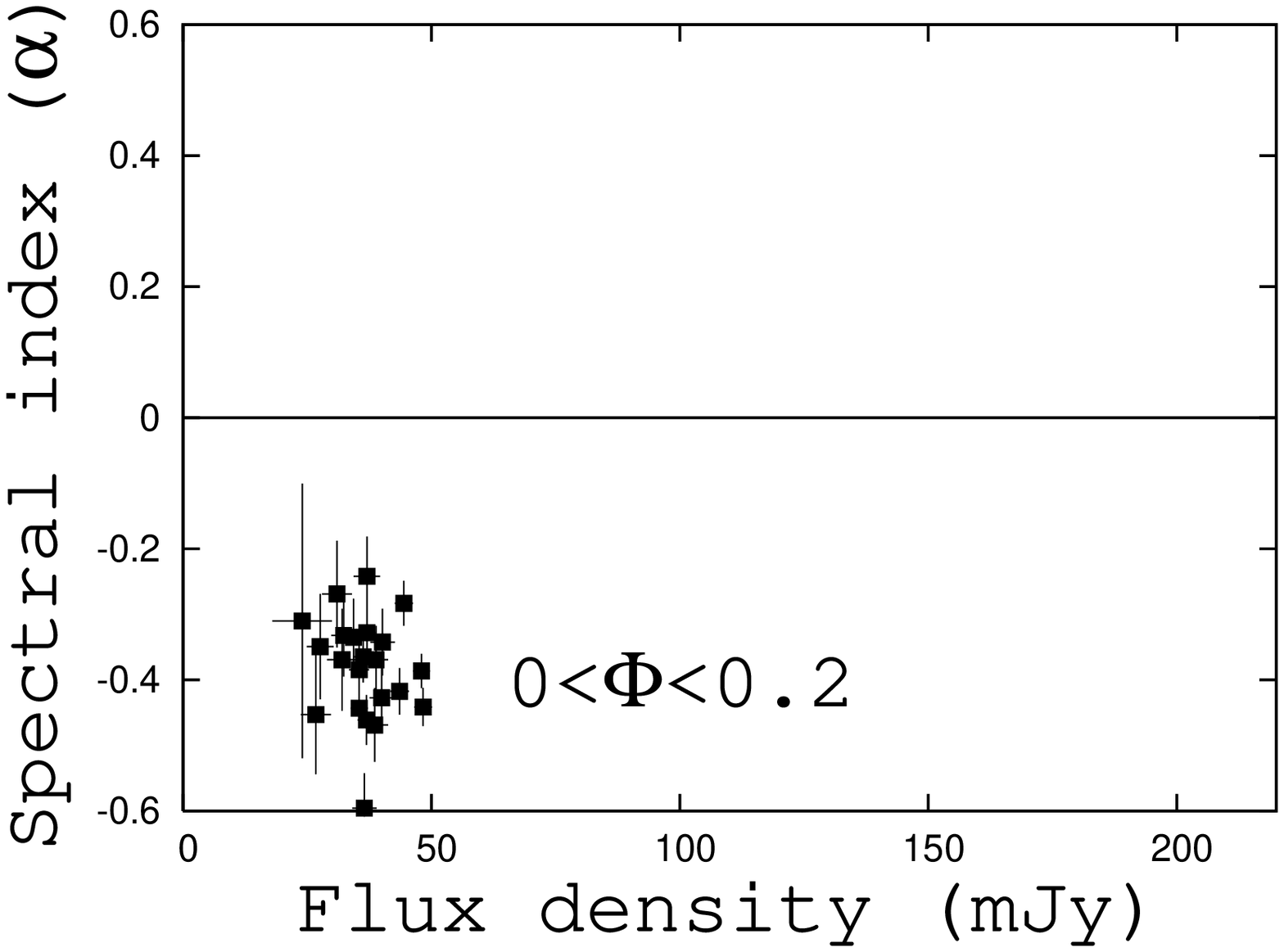}
\includegraphics[width=.17\textheight , angle=-90.]{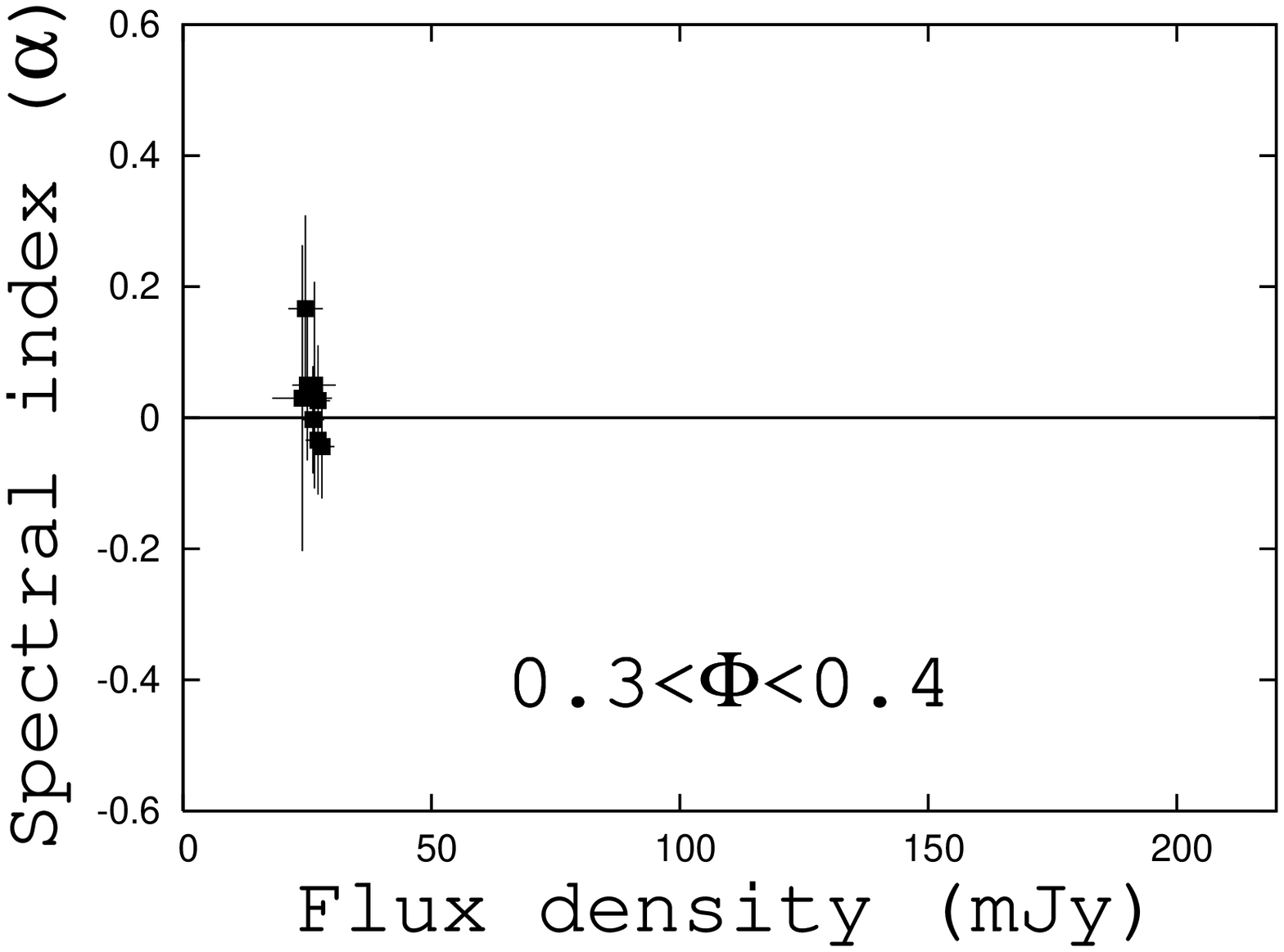}
\includegraphics[width=.17\textheight, angle=-90.]{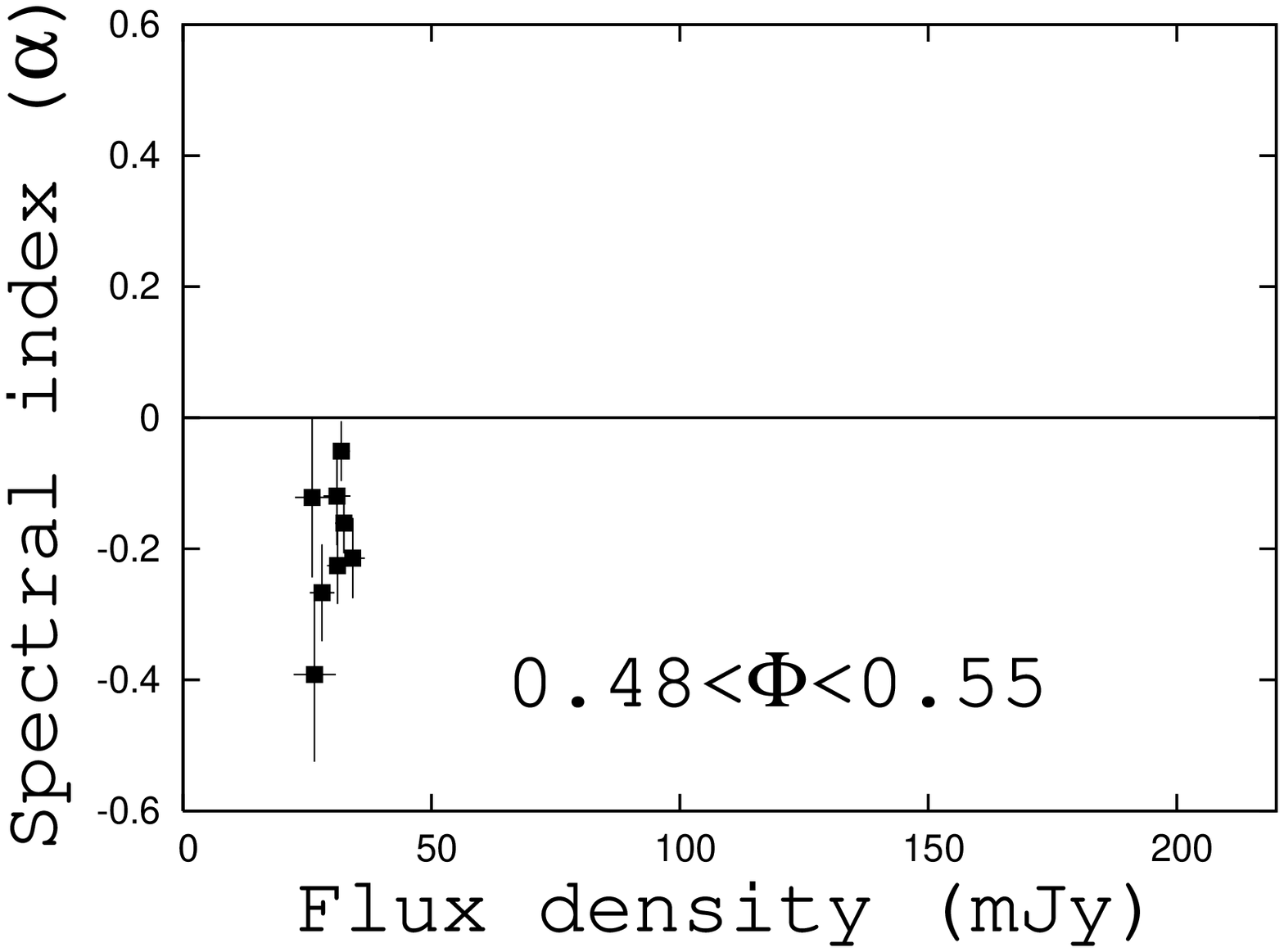}
\includegraphics[width=.17\textheight , angle=-90.]{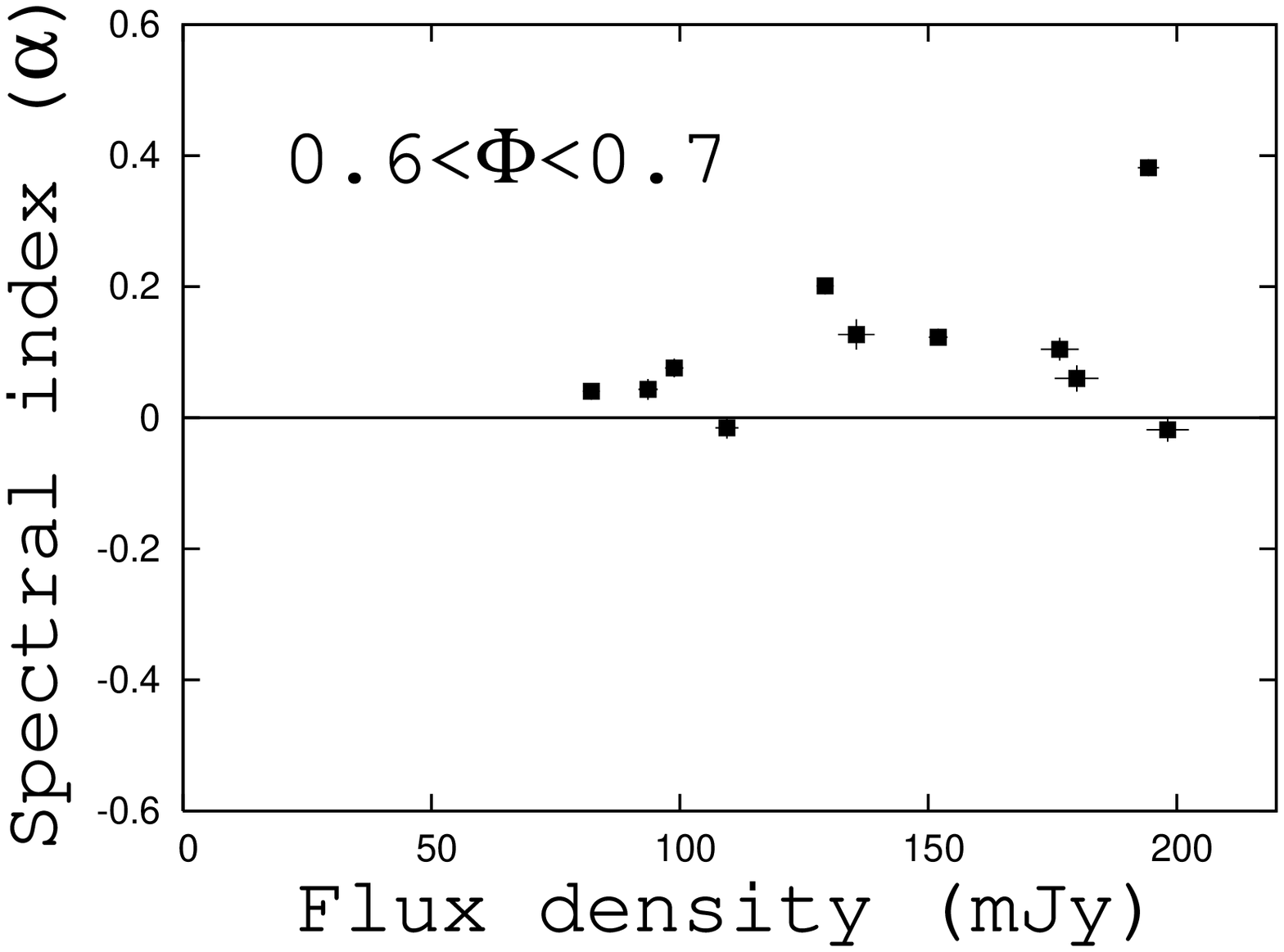}
\includegraphics[width=.17\textheight , angle=-90.]{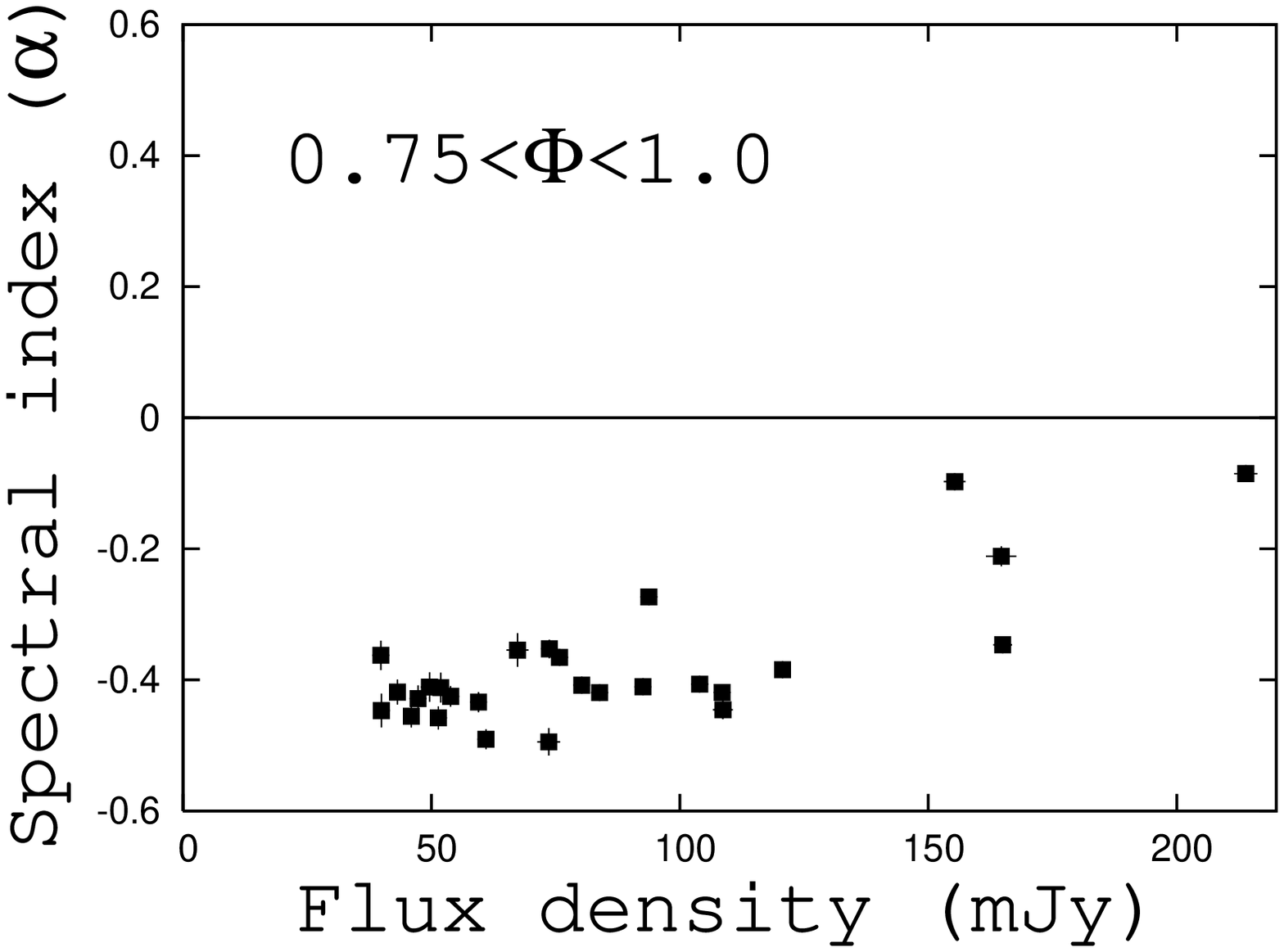}
\end{center}
\caption{
Spectral index vs averaged density flux (8 GHz)  
for the  data set
of Fig. 3. The first plot covers $\Phi$ from 0.0 to 1.0, whereas the other
plots  select different orbital phase intervals. 
The first optically thick interval centered at $\Phi \sim$ 0.35 is associated to low flux density whereas
the second optically thick interval, centered at $\Phi \sim$ 0.65, 
is associated with  high flux density.} 
\label{flal}
\end{figure*}

\begin{figure*}[t!]
 \centering
  \includegraphics[width=.35\textheight , angle=-90.]{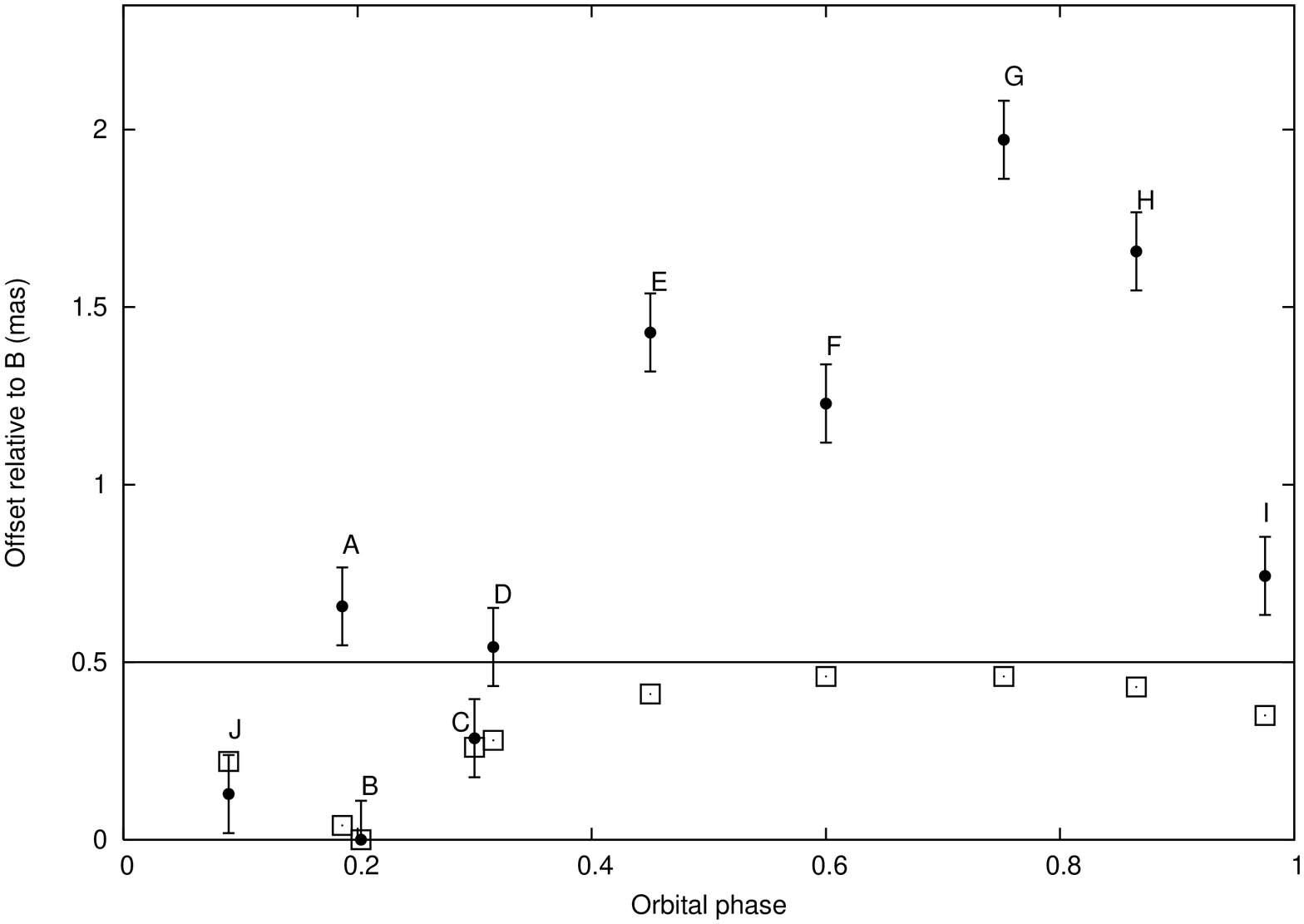}
\includegraphics[width=.35\textheight, angle=-90.]{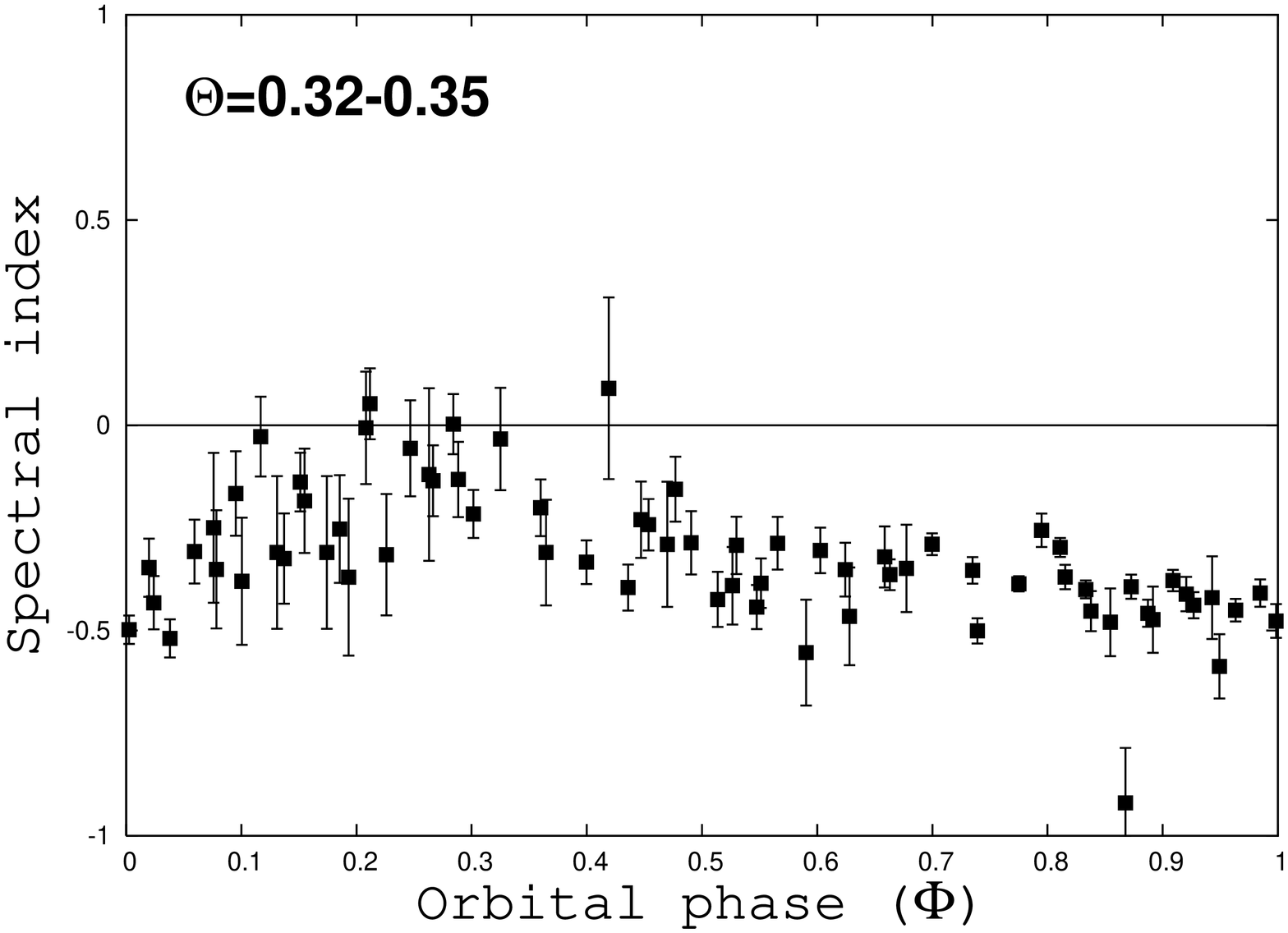}
  \caption{Top: Quantitative comparison of the
astrometry of the radio peaks presented by Dhawan et al. (2006) with respect to the position of the
compact object along the orbit.
Offsets relative to the radio peak B ($\Phi=0.203$) from Dhawan et al. (2006) vs. orbital phase are shown by black dots. The offset is shown with the error bars of 0.1 mas as given by these authors.
The empty squares show the distance in the orbit from phase $\Phi=0.203$ to all the other phases,
 where the VLBA observations have been performed.
These relative distances have been computed  using the orbit from  Casares et al. (2005). The horizontal line at 0.5 mas indicates
the major axis, i.e. the maximum possible distance in the orbit. 
Bottom: spectral
index vs orbital phase (average over $\sim 7$ h) for a subset of the GBI data base at different epochs than  Dhawan
et al. (2006) observations, but covering the same phase ($\Theta$=0.32-0.35) with respect to the
long 1667 d cycle.
}
\label{offset}
\end{figure*}

We analyzed  the radio spectral index using 6.7 years of GBI radio data of \lsp.
We first compared  the radio results with the  predictions of two existing alternative models, then with astrometric results
and finally  with available X-ray and gamma-ray observations.
Our main conclusions are:

\begin{enumerate}
\item  
The  periodic ($P_1=26.5$d) large radio outburst of \lsi  indeed consists
 of two successive outbursts, 
one optically thick and the other optically thin (Figs. 3 and 4).
This peculiar trend does not coincide with an expanding synchrotron-emitting region
producing  a simple spectrum moving
 to lower flux densities and lower frequencies.
 Assuming a prolonged injection of energetic particles one can reproduce the  
 optically thin outburst, but only this one.
The fact that a prolonged injection of energetic particles maintains an outburst
optically thin has been proved for the pulsar PSR B1259-63, but with the
difference to  \lsp, that  the outburst in PSR B1259-63 remains optically thin throughout.
That cannot be compared with the complex sequence (optically thick, optically
thin spectrum) of  \lsp.
This complex sequence finds a natural explanation in the shock-in-jet model:
first there is a continuous outflow with flat or inverted spectrum,
then an event  associated with the first outburst and related to a transition of the accretion disc,
triggers a shock in the   slow, pre-existing optically thick outflow.
At this point  the young growing shock creates
an  optically thin outburst, as it is in fact observed in 
the microquasar GRO J1655-40  and here in \lsp.
\item
During the maximum of the long 1667~d cycle ($\Theta \simeq 1.0$), 
we observe that along the 26.5~d orbit 
the evolution  from an optically thick to an optically thin spectrum
occurs twice, giving
the  $\alpha$ vs $\Phi$ curve  a double-peaked  shape
(Figs.~\ref{oo4} and ~\ref{oo}).
In particular,  the spectrum is optically thick in the interval of 
orbital phase from 
0.22$\pm$0.08 to 0.33$\pm$0.12  and again from 0.50$\pm$0.09 to 0.70$\pm$0.13. 
Therefore, during the orbit a transient jet  occurs twice, each time
preceded by  a steady jet.
This result agrees,  qualitatively and quantitatively with the predictions of 
the  two peak accretion/ejection model,
with the results of three-dimensional
dynamical simulations 
 and finally with  gamma-ray data.
All these results indicate  a scenario with a first ejection   around the
 periastron passage ($\Phi=0.23$) with low radio emission, but high energy emission
because of inverse Compton losses caused by the proximity of the B0 star,
and  a second ejection, delayed $\Delta \Phi=0.3$ (i.e. almost 8 d)
and therefore farther away from the Be star, with negligible losses and  
well observable at radio wavelengths.

\item
We show that the VSOP observations of Taylor et al. (2000) 
occurred within the second optically thick interval. 
That implies that the set of VSOP images 
used to estimate the expansion  velocity show only a core-centered slow
 steady jet and  not a transient jet.

\item
The observations of Dhawan et al. (2006) with some orbit-centered 
features along with orbit-displaced  features,
have been performed at $\Theta=0.32-0.35$.
At that $\Theta$ phase   
there exists only the first optically thick interval, i.e. that
centered around periastron. 
It is in this interval that we find the orbit-centered radio structures (i.e. core-centered steady jet).
In contrast,   the orbital range with only optically thin emission 
is coincident with the orbital range in which there are radio stuctures displaced from the orbit (i.e  shocked ``plasmoids").

\item
When \lsi was detected with MAGIC (mainly for $\Phi=0.5-0.8$),
the energy spectrum was always well fitted by a power law with a photon index
 $\Gamma\simeq 2.6$
that seemed to be  independent of  changes in the  flux level. 
EGRET observations at  $\Theta$=0.4, like MAGIC observations, show  a second
broad  peak at $\Phi\simeq 0.5$, whereas EGRET observations at $\Theta$=0.18
show only a  peak  at periastron passage.
INTEGRAL observations 
give $\Gamma=3.6^{+1.6}_{-1.1}$ where we find $\alpha <0$ 
and a steep power-law  is expected.
In particular, the radio spectral index, $\alpha\simeq -0.5$, we
determined for \lsi, 
corresponds to an electron power law index 
$p \simeq 2$ and  implies high energy emission, by inverse Compton process, 
with a  photon index  $\Gamma >$2.2.

\item
Where we have $\alpha\geq 0$, INTEGRAL observations
give $\Gamma=1.4-1.7$, as is expected for the low/hard X-ray state.
Outside the major  outburst window, the X-ray and radio luminosities of \lsi 
fulfill the 
universal relationship  existing between $L_{\rm X}$ and $L_{\rm R}$ in the low/hard X-ray state.
The low X-ray luminosity and  
the  well established presence of optical emission, related to the
orbital period and to the orbital occurrence of the radio emission,
strongly indicate that \lsi is between a low/hard and a  quiescent state.
 
\end{enumerate}

On the basis of these results we conclude that the object \lsi in a nearly  quiescent
X-ray state - with $\Gamma \simeq  1.4-1.7$ and  with a  core centered slow   jet
with radio spectral index $\alpha \geq $ 0 - 
evolves twice along the 26.5~d orbit  into a steep power-law X-ray state. 
This last state is related to  
an unbroken steep power-law with  $\Gamma \simeq 2.5$  and to a transient jet with radio spectral
index  $\alpha <$0.
The length of the quiescent state phase strongly depends upon 
the 1667~d cycle. 

Our work indicates that the radio spectral index, $\alpha$,
is a powerful tool to distinguish between classes of ejection
and also that a periodic radio source, such as \lsp,  
represents an unique possibility 
for future investigations on the physics of the shock associated to the transient jet.
Concerning the relation
of gamma-ray emission and transient jet,
the first phase of the shock, the growth stage, clearly is of particular interest,
where the dominant cooling mechanism is inverse Compton losses. Future observations
should confirm whether  these Compton losses are responsible for the   unbroken
steep power law. 

\acknowledgments
We are grateful to 
Russ Taylor, Alan Roy,
Karl M. Menten, J\"urgen Neidh\"ofer, and Simon Vidrih
for their  comments.
The Green Bank Interferometer is a facility of the National
Science Foundation operated by the NRAO in support of
NASA High Energy Astrophysics programs.
  

\clearpage

\clearpage

\clearpage

\clearpage

\clearpage

\clearpage

\clearpage
\end{document}